\documentclass[showpacs,prc,showkeys,reprint,aps,floatfix]{revtex4-2}
\usepackage{amsmath}
\usepackage{bm}
\usepackage{gensymb}
\usepackage{graphicx}
\usepackage{mathrsfs}
\usepackage{multirow} 
\usepackage{subfigure}
\usepackage{textcomp}
\usepackage{units}
\usepackage{natbib}
\usepackage[utf8]{inputenc}
\usepackage{cancel}
\newcommand{\eq}[1]{Eq.~(\ref{#1})}
\newcommand{\eqs}[2]{Eqs.~(\ref{#1}) and (\ref{#2})}
\newcommand{\etal}{\emph{et al.~}}

\newcommand{\he}{$^3$He}
\newcommand{\nhe}{n-$^3$He}
\renewcommand{\vec}[1]{\bm{#1}}
\newcommand{\werb}[3]{\unit[(#1 $\pm$ #2)]{#3}}  
\newcommand{\WERB}[6]{\unit[(#1 $\pm$ #2 (#3) $\pm$ #4 (#5)\,)]{#6}} 
\newcommand{\abs}[1]{\left\vert #1 \right\vert}
\newcommand{\ComDb}{$-5.411$}
\newcommand{\ComDbstat}{$0.031$}
\newcommand{\ComDbsys}{$0.039$}

\newcommand{\OLDDb}{$-5.610$}
\newcommand{\OLDDbstat}{$0.027$}
\newcommand{\OLDDbsys}{$0.032$}

\DeclareMathOperator{\sech}{sech}

\begin{document}

\title{Neutron interferometric measurement of the scattering length \\difference between the triplet and singlet states of \nhe\ }
\author{M. G. Huber}
\email{michael.huber@nist.gov}
\author{M. Arif}
\author{W. C. Chen}
\altaffiliation[Also at ]{University of Maryland, College Park, MD 20742}
\author{T. R. Gentile}
\author{D. S. Hussey}
\affiliation{National Institute of Standards and Technology, Gaithersburg, MD 20899, USA}
\author{T. C. Black}
\affiliation{University of North Carolina-Wilmington, Wilmington, NC 28403, USA}
\author{D. A. Pushin}
\altaffiliation[Also at ]{Institute for Quantum Computing, Waterloo, ON N2L 3G1, Canada}
\affiliation{University of Waterloo, Waterloo, ON N2L 3G1, Canada}
\author{C. B. Shahi}
\author{F. E. Wietfeldt}
\affiliation{Tulane University, New Orleans, LA 70188, USA}
\author{L. Yang}
\affiliation{University of Illinois at Urbana-Champaign,  Urbana, IL 61801, USA}\date{\today}
\begin{abstract}
We report a determination of the \nhe\ scattering length difference $\Delta b^{\prime} = b_{1}^{\prime}-b_{0}^{\prime} =  $
\WERB{\ComDb}{\ComDbstat}{statistical}{\ComDbsys}{systematic}{fm} between the triplet and singlet states using a neutron interferometer.
This revises our previous result  $\Delta b^{\prime} = $ \WERB{\OLDDb}{\OLDDbstat}{statistical}{\OLDDbsys}{systematic}{fm} obtained using the same technique in 2008 [M.\ G.\ Huber, Phys.\ Rev.\ Lett.\
\textbf{20}, 102 (2009) \& M.\  G.\ Huber, Phys.\  Rev.\ Lett.\ \textbf{17}, 103 (2009)]. 
This revision is due to a re-analysis of the 2008 experiment that now includes a systematic correction caused by magnetic field gradients near the  \he\ cell which had been previously underestimated.
Furthermore, we  more than doubled our original data set from 2008 by acquiring six months of additional data in 2013.
Both the new data set and a re-analysis of the older data are in good agreement. 
Scattering lengths of low Z isotopes are valued for use in few-body nuclear effective field theories,  provide important tests of modern nuclear potential models and in the case of \he\ aid in the interpretation of neutron scattering from quantum liquids.
The difference $\Delta b^{\prime}$  was determined by measuring the relative phase shift between two incident neutron polarizations caused by the spin-dependent interaction with a polarized \he\ target. 
The target \he\ gas was sealed inside a small, flat windowed glass cell that was placed in one beam path of the interferometer. 
The relaxation of \he\ polarization  was monitored continuously with neutron transmission  measurements.
The neutron polarization and spin flipper efficiency were determined separately using  \he\ analyzers and two different polarimetry analysis methods.  
A summary of  the measured scattering lengths for \nhe\ with a comparison to nucleon interaction models is given.
\end{abstract}
\maketitle

\section{Introduction  \label{Sec_Intro}}
Understanding the properties of nuclei from the point of view of a collection of individual interacting nucleons is an important goal of nuclear physics \cite{lazauskas_2005_PhysRev,Viviani_2005_NuclPhys}. 
Unfortunately, direct calculations of  low-energy nuclear phenomena using Quantum Chromodynamics (QCD) is currently impractical.
Instead properties of few-body nuclear systems are calculated using a variety of phenomenological models.
The prevailing  two nucleon (NN) models, AV18 \cite{Wiringa_1995_PhysRev}, CD-Bonn \cite{Machleidt_2001_PhysRev}, and Nijmegen \cite{Stoks_1994_PhysRev}, incorporate a fit to \emph{np} and \emph{pp} scattering  data \cite{Stoks_1993_PhysRev} for  energies up to \unit[350]{MeV}.
Problems arise with NN models when applying them to systems containing more than two nucleons. 
This is most evident by their failure to accurately predict the binding energy of the triton by \unit[800]{keV} \cite{Pieper_2001_PhysRev}.
For this reason three nucleon interactions (3N), which arise in lowest order in the meson exchange model from the exchange of two pions between three separate nucleons, are included with NN models to describe larger few-body systems.  
Three nucleon interactions, including Urbana \cite{Pudliner_1997_PhysRev}, Tucson Melbourne \cite{Coon_1979_NuclPhys}, Brazil \cite{Robilotta_1986_NuclPhys} or Illinois \cite{Pieper_2001_PhysRev} potentials, can correct for this and now accurately predict many nuclear levels for atomic number up to $13$ \cite{Wiringa_2014_PRC,Wiringa_2002_PRL,Pieper_2001_PhysRev}.  
However, the increase in the prediction accuracy of binding energies has not meant that NN+3N models have accurately predicted low-energy scattering data in systems involving more than two nucleons \cite{Viviani_2001_PRL,Black_2003_PRL}.
\par
Nuclear effective field theories\cite{Epelbaum_2002_PhysRev, Entem_2003_PhysRev}  have also been a successful approach to understand few-body nuclear systems.
Effective field theories separate nucleon interactions into two distinct energy regions that are separated by the pion mass. 
Below the pion mass threshold interaction diagrams are explicitly calculated.
For higher-energy processes, the interactions are described by using low-energy  observables such as scattering lengths to parameterize mean-field behavior. 
For instance using the triton binding energy Kirscher \etal\cite{Kirscher_2009_Archive} predicted a value of  the scattering length for the singlet state  in \nhe\  to \unit[8]{\%} relative uncertainty.
Although presently not as precise as other approaches, effective field theories are attractive because they provide clear theoretical uncertainties from estimates of the relative contribution of higher order terms \cite{Bedaque_1999_NuclPhys}.
\par 
A final motivation for measuring the \nhe\ scattering length to high precision is that it also arises in the study of quantum liquids. 
Inelastic neutron scattering in liquid \he\ for a momentum $\vec{Q}$ and energy transfer $E$  is described by a dynamic structure factor $S(\vec{Q},E)$ \cite{Glyde_1995_book}. 
The dynamic structure factor can be separated into coherent $S_c$ and incoherent $S_i$ terms as \cite{Skold_1976_PRL,Glyde_2000_PhysRev}
\begin{eqnarray}
S(\vec{Q},E)= S_c(\vec{Q},E)+\frac{\abs{\sqrt{3}\Delta b^\prime}^2}{\abs{b^\prime_{0}+3b^\prime_{1}}^2}S_i(\vec{Q},E) \label{eq1}.
\end{eqnarray} 
One can see that the incoherent density term is weighted by a ratio composed of both the triplet $b^\prime_{1}$ and singlet  $b^\prime_{0}$ scattering lengths where $\Delta b^{\prime} = b_{1}^{\prime}-b_{0}^{\prime}$.
(The real part of the scattering length is denoted by $^\prime$). 
For the \nhe\ system  the sum  $b^\prime_{0}+3b^\prime_{1}$  has been previously measured to $ < $ \unit[1]{\%} relative uncertainty.
An accurate determination of $S_c(\vec{Q},E)$ and $S_i(\vec{Q},E)$ from $S(\vec{Q},E)$ relies on determining  $\Delta b^{\prime}$ to similar precision. 
\par
Neutron scattering lengths can be determined very precisely using neutron interferometry.  
In the case of silicon, neutron interferometry has been utilized to measure the scattering length to \unit[0.005]{\%} relative uncertainty \cite{Ioffe_1998_PhysRev}. 
In the last few years, there have been several precision measurements using neutron interferometry with light nuclei targets.  
These include measurements of the spin-independent scattering length $b^{\prime}$ of n-$^1$H, n-$^2$H \cite{Schoen_2003_PhysRev}, and n-$^3$He \cite{Huffman_2004_PhysRev,Ketter_2006_EPJ} to less than one percent relative uncertainty.  
\par
Here we report a determination of the scattering length difference $\Delta b^{\prime} = b_{1}^{\prime}-b_{0}^{\prime}$ of \nhe\ using a neutron interferometer that is based on a re-analysis of  Huber \etal \cite{Huber_2009_PRLa,Huber_2009_PRL} and additional data. 
This effort was revisited in order to quantify the phase shift due to a non-uniform magnetic field near the target cell.
Phase shifts caused by a non-uniform magnetic field were previously underestimated and only  included in our analysis as an overall systematic uncertainty.
In this work we have more thoroughly estimated the magnetic field gradient induced phase shift and now include an additional correction for this effect.
\par
A review of the relevant neutron optical theory is covered in Sec.~\ref{Sec_Theo}.  
The experimental setup and measured phase shift caused by the \he\ target sample  is discussed in Sec.~\ref{Sec_Proc}. 
A discussion of the source of correction to the 2008 data set can be found in Sec.~\ref{sec_field}.
In Sec.~\ref{Sec_Aux} we describe neutron polarimetry measurements using \he\ analyzers with two different analysis methods.  
Systematically limiting this technique is the uncertainty in the triplet absorption cross section of \he. 
This limit in determining $\Delta b^{\prime}$ and other non-statistical uncertainties are discussed in Sec.~\ref{SubSec_Syst}.  
Finally, we compare world averages of the current experimental results of the coherent and incoherent scattering lengths to various theoretical predictions in Sec.~\ref{Sec_Con}. 

\section{Neutron Scattering   \label{Sec_Theo}}
Neutron scattering from a single target atom can be described by the wave function \cite{Goldberger_1947_PhysRev}
\begin{eqnarray}
\Psi=e^{i\vec{k}\cdot\vec{r}}+\frac{e^{ikr}}{r}f(\theta). \label{Eq_Theo_1}
\end{eqnarray}
The first term in \eq{Eq_Theo_1} describes the incident neutron where $\vec{k}$ is the incident wave vector and $\vec{r}$ is the position of the neutron. 
The latter term corresponds to the scattered wave in the first Born approximation with a scattering amplitude $f(\theta)$ that can be expanded in terms of  $k$ as \cite{Sears_1989_Book}
\begin{eqnarray}
f(\theta)=-a+ika^2+\mathcal{O}(k^2) + \ldots \approx -a, \label{Eq_Theo_2}
\end{eqnarray}
where $a$ is called the free scattering length. 
The approximation in \eq{Eq_Theo_2} is valid because the magnitude of $a $ is of $\mathcal{O}($\unit[1]{fm}$)$ and for low energy neutrons  $ k$ is of $\mathcal{O}($\unit[$10^{-4}$]{fm$^{-1}$}$)$.
In general, $a$ is complex such that $a=a^\prime +ia^{\prime\prime}$ where $a^\prime, a^{\prime\prime}$ are both real numbers.  
As discussed later, neutron absorption by the target atom is related to the imaginary term $a^{\prime\prime}$.  
Most importantly $a$  represents a  measurable quantity of the interaction that is unique  for each nuclear  isotope.
In neutron interferometry, even when considering a gas target, the forward scattered momentum transfer is zero.   
For this reason, it is more relevant to define everything using  the bound scattering length 

\begin{eqnarray}
b=\left(\frac{M_\mathrm{N}+m_\mathrm{n}}{M_\mathrm{N}}\right)a. \label{Eq_Theo_3} 
\end{eqnarray}
Here $M_\mathrm{N}$ and $m_\mathrm{n}$ are the mass of the target atom and neutron, respectively.  
\begin{figure}[t]
	\centering
	\includegraphics[width=3.375in]{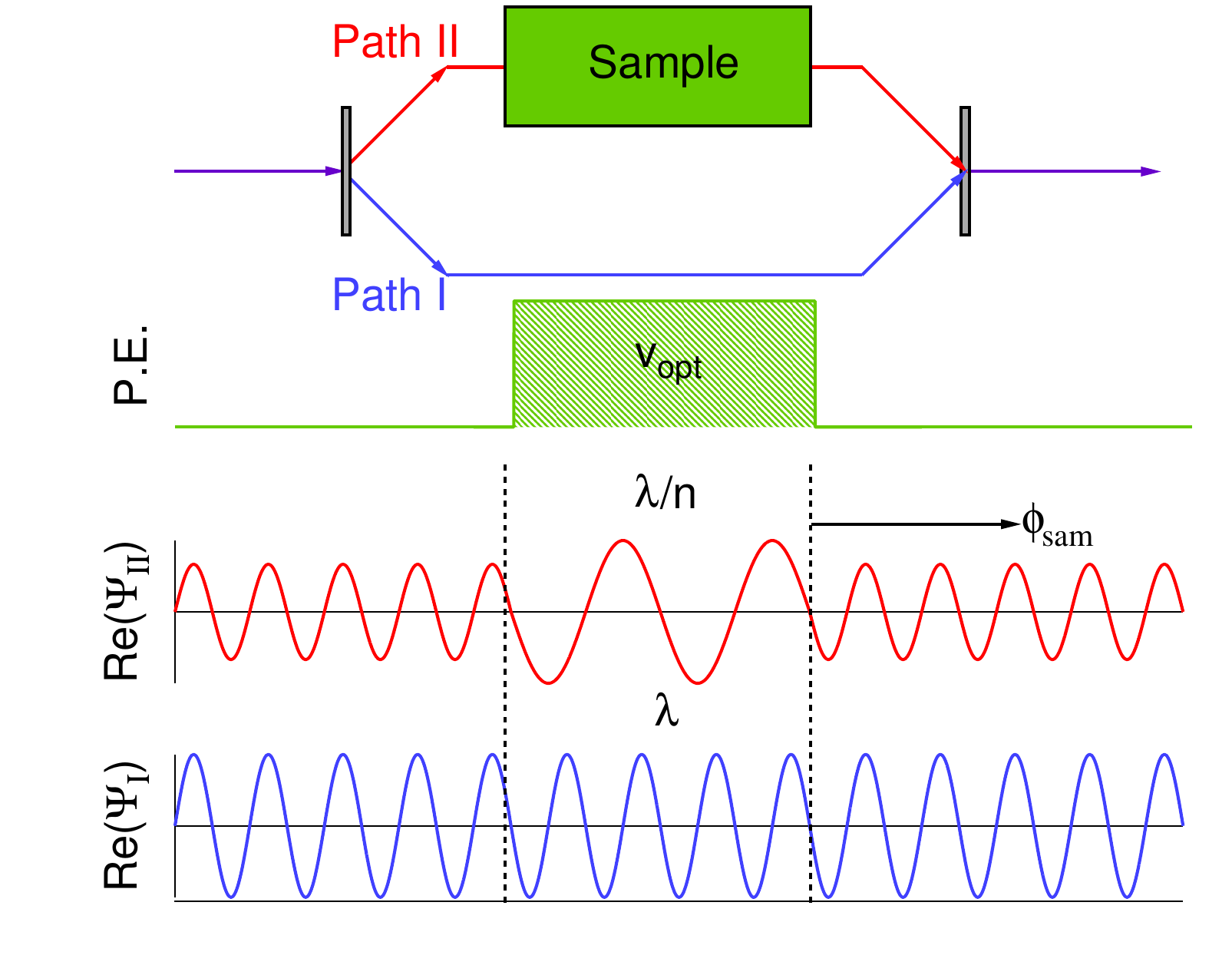}
	\caption{(Color online) A simplified schematic of an interferometer. A beam splitter separates a neutron wave function into two  paths that are recombined in a final analyzing beam splitter before being detected.  A sample placed in one path causes a relative phase shift $\phi_\mathrm{sam}$ because of a change in wavelength due to the index of refraction of the material. }  
	\label{Fig_Theo_1}
\end{figure}   
\par
To describe neutrons scattering from a homogeneous material one uses the time-independent Schr\"{o}dinger equation
\begin{eqnarray}
\left(\frac{\hbar^2K^2}{2m_\mathrm{n}}+ V_\mathrm{opt}\right)\psi =\frac{\hbar^2k^2}{2m_\mathrm{n}}\psi \mbox{,} \label{Eq_Theo_4} 
\end{eqnarray}
where $K$ is the magnitude of the neutron wave vector in the material. 
The optical potential
\begin{eqnarray}
V_\mathrm{opt}= \frac{2\pi\hbar^2}{m_\mathrm{n}}Nb, \label{Eq_Theo_5} 
\end{eqnarray}
is the effective potential of the material with an atom density $N$.
The  index of refraction 
\begin{eqnarray}
n=\frac{K}{k}=\sqrt{1-\frac{\lambda^2 Nb}{\pi}}, \label{Eq_Theo_6} 
\end{eqnarray}
of the material  can be derived using \eqs{Eq_Theo_4}{Eq_Theo_5}.
Here $\lambda=2\pi/k$ is the neutron wavelength in vacuum.
The index  of refraction for neutron optics is conceptually the same as for light optics with two subtle differences.
For one the index of refraction for neutrons is only a small deviation from unity;  ie. $1-n =\mathcal{O}(10^{-6})$  and  generally $n$ is less than one.
\par
A neutron interferometer splits the neutron's wave function along two spatially  separate paths labeled I and II.
When a sample of thickness $D$ is introduced into beam path \rmfamily{II} there is a phase difference relative to the path \rmfamily{I} of
\begin{eqnarray}
\phi_\mathrm{sam} = k(1-n)D = -\lambda NDb^\prime \mbox{.} \label{Eq_Theo_7} 
\end{eqnarray}
This phase shift  is due to the change of the neutron wavelength as it passes through a material.  
A conceptual illustration of this is provided in Figure \ref{Fig_Theo_1}.
A neutron interferometer is extremely sensitive to phase differences between paths and therefore can be used to measure $\phi_\mathrm{sam}$ to high precision.
Along with information of the quantities $\lambda$, $N$ and $D$, often performed using  individual  auxiliary  measurements,  one can use \eq{Eq_Theo_7} to determine the scattering length of the material.
\par
In the case of neutrons with spin $\vec{\sigma}_\mathrm{n}$ incident on a target with nuclear spin $\vec{I}$ the  scattering length can be written as
\begin{eqnarray}
b=b_c+\frac{\displaystyle 2b_i}{\displaystyle \sqrt{I(I+1)}}\vec{\sigma}_\mathrm{n}\cdot\vec{I} \mbox{,} \label{Eq_Theo_8}
\end{eqnarray}
where $b_c$ and $b_i$ are called the coherent and incoherent scattering lengths, respectively. 
Despite its name the incoherent scattering length is due to a coherent interaction and corresponds to the spin-dependent part of the scattering length.
The scattering lengths for a given spin channel, $\vec{J}=\vec{I}\pm\vec{\sigma}_\mathrm{n}$, are defined as
\begin{subequations}\label{eq11}
\begin{eqnarray}
b_{+} =& b_c+\displaystyle \sqrt{\frac{\displaystyle g_-}{\displaystyle g_+}}b_i    \label{Eq_Theo_9} \\
b_{-} =& b_c- \displaystyle \sqrt{\frac{\displaystyle g_+}{\displaystyle g_-}} b_{i}\label{Eq_Theo_10}.
\end{eqnarray}
\end{subequations}
Here $g_+= (I+1)/(2I+1)$ and $g_-=1-g_+=I/(2I+1)$ are statistical weight factors.  
\eqs{Eq_Theo_9}{Eq_Theo_10} are for general systems;
for \nhe\ there is a triplet (J $ =1$) and singlet (J $ =0$) channel allowing us to write the triplet scattering length as $b_{+} \equiv b_{1}$ and the singlet scattering length as $b_{-} \equiv b_{0}$
(likewise, $g_+\rightarrow g_1$ and $g_-\rightarrow g_0$).
Inverting \eqs{Eq_Theo_9}{Eq_Theo_10}, the  \he\ coherent and incoherent scattering lengths become 
\begin{subequations}\label{eq12}
\begin{eqnarray}
b_c & = & g_1 b_1 + g_0 b_0  \label{Eq_Theo_11} \\ 
b_i & = & \sqrt{g_1 g_0}(b_1 -b_0) \label{Eq_Theo_12}.
\end{eqnarray}
\end{subequations}
\par
The total cross section for the \nhe\ interaction is $\sigma_t=\sigma_s+\sigma_a$.  
Here $\sigma_s$ is the scattering cross section given by
\begin{eqnarray}
\sigma_s= \sigma_{s,c}+\sigma_{s,i} = 4\pi\left|b_c\right|^2+4\pi\left|b_i\right|^2. \label{Eq_Theo_13}
\end{eqnarray}
The absorption cross section $\sigma_a$ is related to the imaginary part of the scattering length $b^{\prime\prime}$ by the optical theorem \cite{Feenberg_1932_PhysRev}
\begin{eqnarray}
\sigma_a= \frac{4\pi}{k}b^{\prime\prime}. \label{Eq_Theo_14}
\end{eqnarray}
The measured unpolarized neutron absorption cross section for \he(\emph{n},\emph{p})$^3$H is $\sigma_\mathrm{un} =$\werb{5333}{7}{b} \cite{Mughabghab_2006_Book} at the reference thermal  neutron wavelength $\lambda_\mathrm{th} = $ \unit[1.798]{\AA}.
The uncertainty quoted for  $\sigma_\mathrm{un}$, as well as all other uncertainties quoted below, is a standard uncertainty with a confidence level of  \unit[68]{\%}. 
For \he(\emph{n},$\gamma$)$^4$He the absorption cross section is \werb{54}{6}{\micro b} \cite{Wolfs_1989_PRL} at these energies and
thus for our purposes  can be ignored.  
\par
The absorption cross section for polarized neutrons incident on polarized \he\ nuclei is given by
\begin{eqnarray}
\sigma_a=\sigma_\mathrm{un} \mp P_3\sigma_p, \label{Eq_Theo_15}
\end{eqnarray}
where $P_3$ is the \he\ polarization.  
The $\mp$ sign in \eq{Eq_Theo_15} represents the cases where the neutron and \he\ spins are aligned parallel $(-)$  or anti-parallel $(+)$. 
Similar to \eqs{Eq_Theo_11}{Eq_Theo_12}, $\sigma_\mathrm{un} $ and $\sigma_p$ can be defined in terms of singlet and triplet contributions as 
\begin{subequations}
\begin{eqnarray}
\sigma_\mathrm{un}= g_1\sigma_1+g_0\sigma_0 \label{Eq_Theo_16}\\ 
\sigma_p=g_0\left(\sigma_0-\sigma_1 \right). \label{Eq_Theo_17}
\end{eqnarray}
\end{subequations}
Since  $\sigma_a$ is dominated by the singlet channel it is often assumed that $\sigma_1=0$ so that $\sigma_\mathrm{un}=\sigma_p$. 
Although $\sigma_1$ is small there is no theoretical justification for assuming $\sigma_1$ to be precisely zero. 
Neutron transmission measurements are consistent with $\sigma_1=0$ only at the level of a few percent.  
The uncertainty in the triplet absorption cross section leads to the largest systematic uncertainty in both neutron interferometric and pseudomagnetic spin precision measurements of $\Delta b^\prime$.
This is discussed in more detail in Section \ref{Sec_Abs}.   
\section{Experimental Procedure \label{Sec_Proc}}

\begin{figure*}[t!]
    \centering
    \begin{subfigure}
		 \centering
        \includegraphics[width=3.375in]{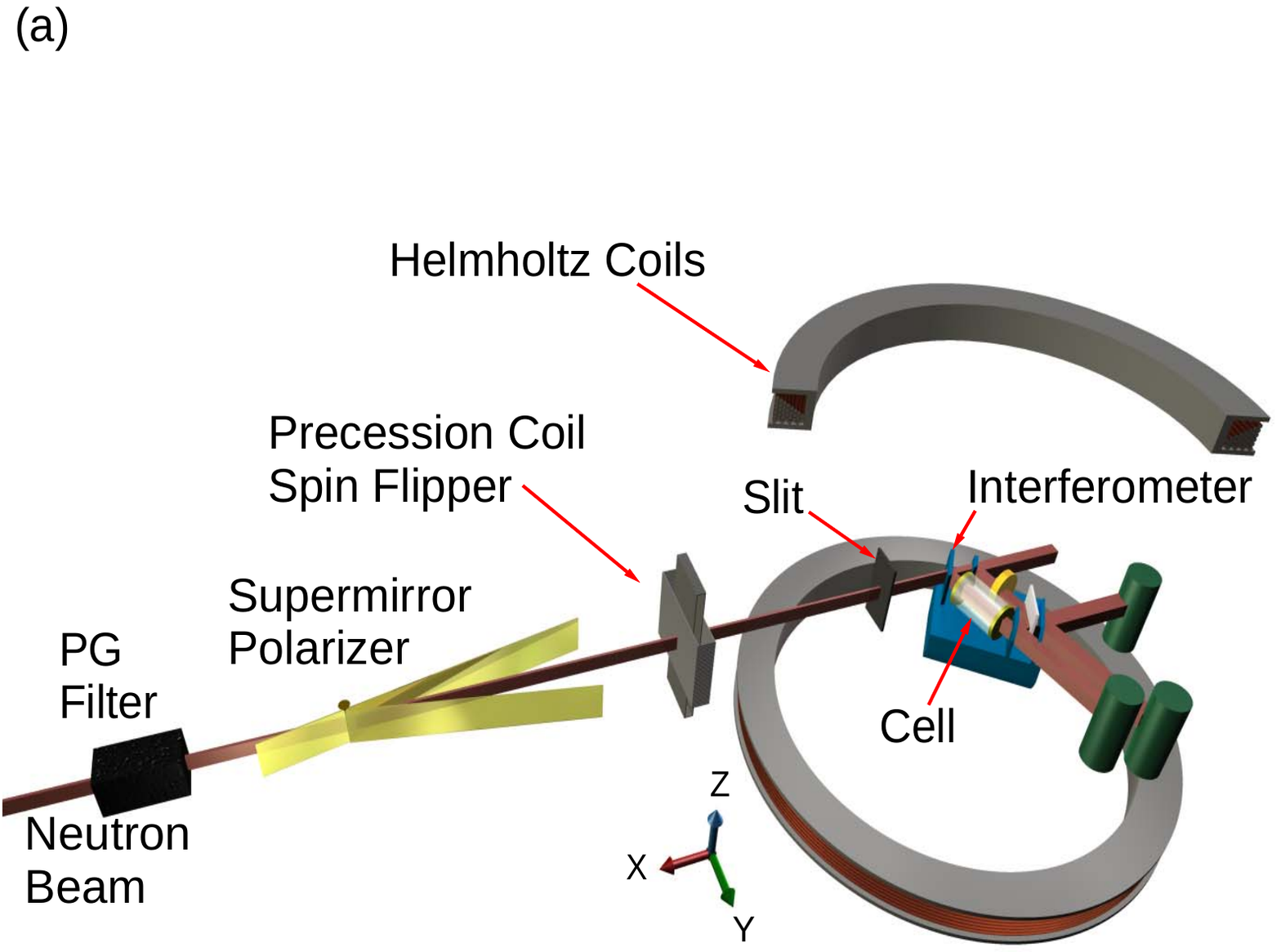}
    \end{subfigure}%
		~
    \begin{subfigure}
		 \centering
        \includegraphics[width=3.375in]{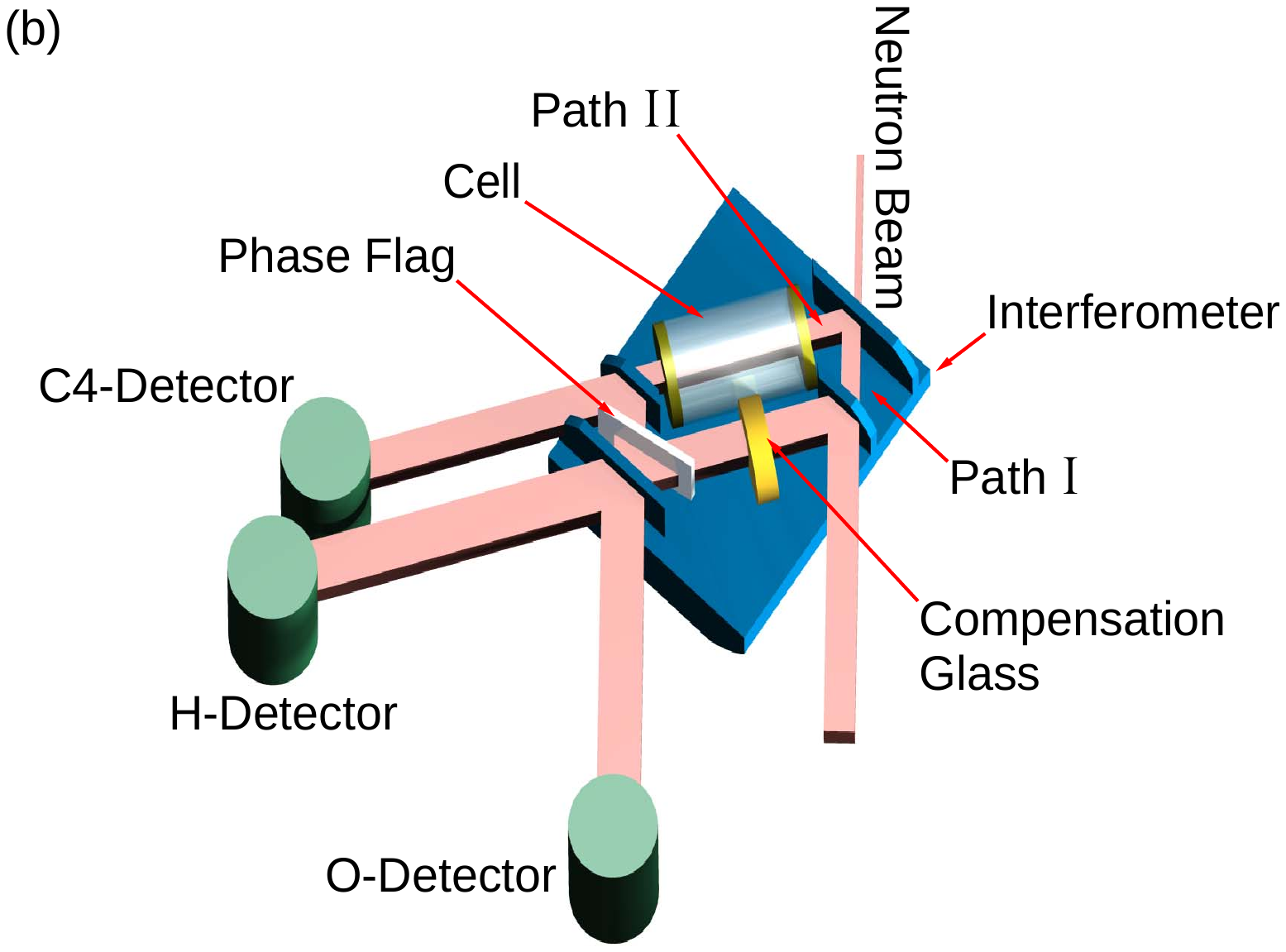}\label{Fig_Proc_1b}
    \end{subfigure}
    \caption{(Color online) The \nhe\  experiment (Not to Scale). 
	(a) A monochromatic neutron beam entering from the left is polarized by a supermirror.   The polarization direction can be changed using a precession coil spin flipper. 
	A \unit[2]{mm} $\times$ \unit[6]{mm} slit provided collimation just before the interferometer. 	
	(b) Neutrons  Bragg diffract in the first blade of the 		interferometer  coherently splitting the neutron into two separate paths. 
  The two paths are diffracted in separate mirror blades so that they mix and interfere at the analyzer blade of  the interferometer.  
	One neutron path contains the \he\ target cell while the other path contains \unit[8]{mm} of boron-free glass to compensate for 	
	the phase shift caused by the target cell windows. 
	A quartz phase flag is rotated to vary the intensity in the two \he\ filled proportional counters labeled the O- and H-beam 	
	detectors. 
	The \he\ polarization was monitored by a third \he\ detector labeled C4.}\label{Fig_Proc_1}
\end{figure*}
We used a neutron interferometer to measure the phase difference between two polarized neutron states that are transmitted through a polarized \he\ target cell.
Neutrons are polarized  in the vertical direction and can be flipped by $\pi$ radians using a precession coil spin flipper.
The target \he\ polarization direction stays constant throughout a measurement, but its magnitude  decreases exponentially in time.
For neutrons with spin state $\uparrow$ and $\downarrow$ incident on a polarized \he\ sample we can insert the effective scattering length for \he\ (\eq{Eq_Theo_8})  
\begin{eqnarray}
b^\prime=b_c^\prime\pm\frac{\displaystyle b_i^\prime}{\displaystyle \sqrt{3}}, \label{Eq_Proc_1a} 
\end{eqnarray}
into \eq{Eq_Theo_7}  and find the phase difference
\begin{align}
\Delta\phi_0 &=\phi_\mathrm{sam}^{\uparrow}-\phi_\mathrm{sam}^{\downarrow} \nonumber\\
&=-\lambda N_3D_3\bigg[\bigg(b_c^\prime + N_+\frac{\displaystyle b_i^\prime}{\displaystyle \sqrt{3}} -N_-\frac{\displaystyle b_i^\prime}{\displaystyle \sqrt{3}}\bigg)  \nonumber\\
&\quad -\bigg(b_c^\prime - N_+\frac{\displaystyle b_i^\prime}{\displaystyle \sqrt{3}} +N_-\frac{\displaystyle b_i^\prime}{\displaystyle \sqrt{3}}\bigg) \bigg] \label{Eq_Proc_1b} 
\end{align} 
where $N_{3}$ is the \he\ number density and $D_{3}$ is the active length of the target cell. 
Here $N_\pm =(1\pm P_3)/2$ is the fraction of \he\ nuclei in each polarization state.  
Using \eqs{Eq_Theo_12}{Eq_Proc_1b} one finds that the  phase difference  between opposite neutron spin states is related to the triplet and singlet scattering lengths by
\begin{eqnarray}
b_1^\prime -b_0^\prime =\frac{\displaystyle -2\Delta\phi_0}{N_{3}\lambda D_{3} P_{3}}, \label{Eq_Proc_1} 
\end{eqnarray}
The factors in the denominator of \eq{Eq_Proc_1} are determined simultaneously  with $\Delta\phi_0$ by measuring the spin-dependent transmission of neutrons through the \he\ cell. 
This is an advantage over typical interferometric measurements of $b^\prime$ in that none of the factors in the denominator need to be determined individually; the spin-dependent transmission asymmetry determines their product directly.

\subsection{Neutron Interferometer and Facility \label{SubSec_NIOF}}
This experiment was done at the National Institute of Standards and Technology's (NIST) Center for Neutron Research (NCNR) in Gaithersburg, MD.  
A \unit[20]{MW} reactor produces a steady source of neutrons that are moderated by a liquid hydrogen cold source \cite{Williams_2002_Physica}.
\begin{figure}[t]
	\centering
	\includegraphics[trim=5cm 1cm 1cm 1cm, clip=true,width=3.375in]{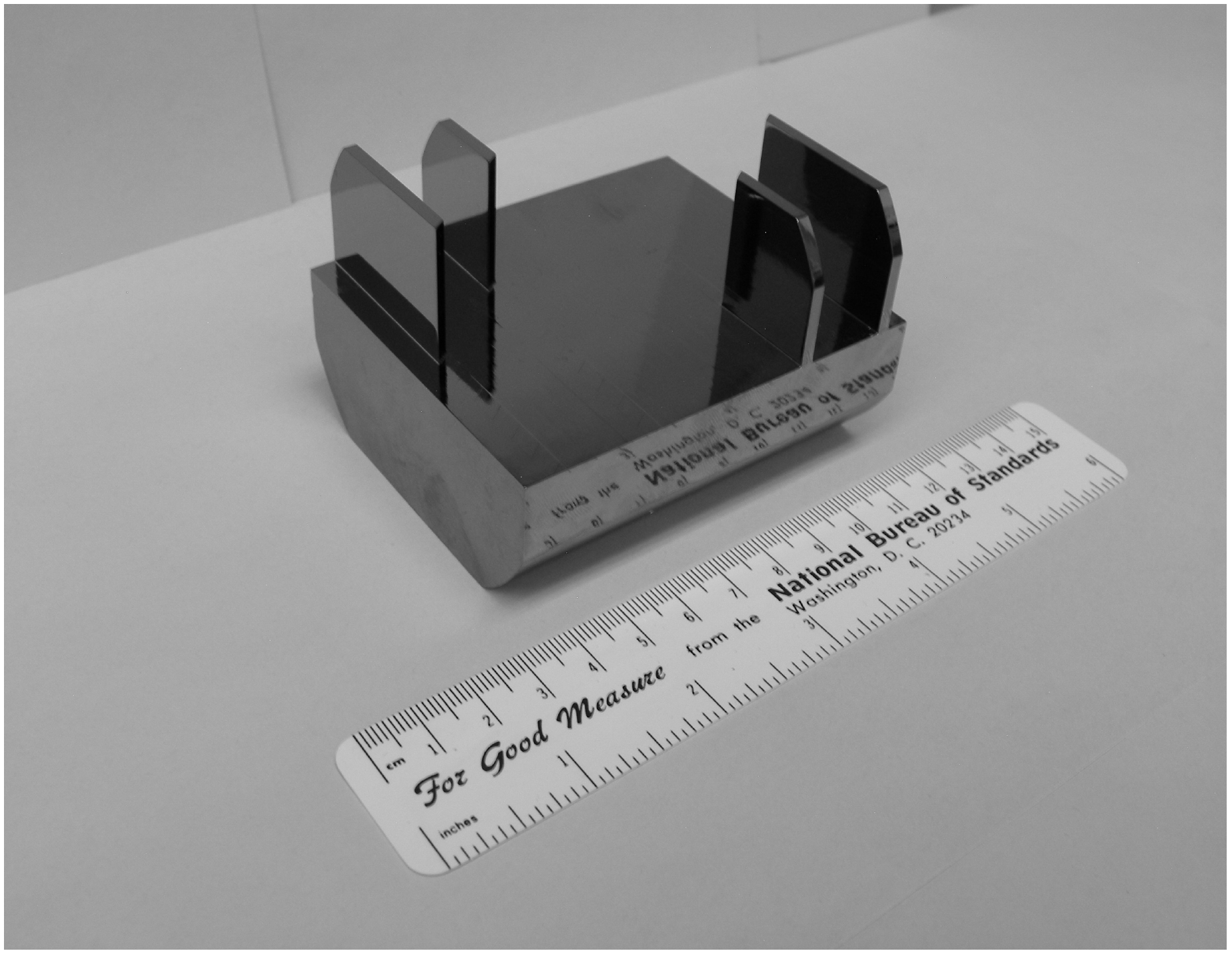}
	\caption{The skew symmetric interferometer made by Cliff Holmes and others at the University of Missouri-Columbia machine shop (on loan from   
	Dr.~Samuel A.~Werner).}
	\label{Fig_Proc_2}
\end{figure}
These moderated neutrons are transported from the cold source to several neutron instruments by $^{58}$Ni coated glass guides. 
At the Neutron Interferometer and Optics Facility (NIOF) a double  monochromator assembly reflects \unit[2.35]{\AA} neutrons into an environmentally  isolated enclosure \cite{Arif_1994_VibrMonitCont}.  
The first monochromator is a single pyrolytic graphite PG(002) crystal that reflects neutrons out of the neutron guide and toward a second monochromator \unit[3]{m} away.  
This second monochromator vertically focuses the beam using nine individually adjustable \unit[1]{cm} x \unit[5]{cm} PG(002) crystals \cite{Pushin_2008_PRL}.  
Further details of the facility can be found in \cite{M.G.Huber_2014Inpreparation_NI&MiPRSAASDaAE}.
\par
A schematic of the experiment inside the enclosure is shown in Figure \ref{Fig_Proc_1}.  
After the double monochromator assembly a pyrolytic graphite filter was used to remove $\lambda/2= $ \unit[1.175]{\AA} neutrons from the beam.  
Neutrons passing through the filter were polarized by a transmission-mode supermirror polarizer \cite{Ioffe_1996_JPhysSocJap}.  
The polarizer consisted of two separate {\unit[0.64]{m} and \unit[0.73]{m}} long supermirrors that were slightly offset so that no incoming neutrons had clear line of sight to the interferometer. 
Neutrons in the incorrect spin state were reflected from the supermirror and absorbed on cadmium shielding. 
\par
Immediately downstream of the supermirror polarizer was a precession coil spin flipper made from two orthogonal aluminum wire coils \cite{Badurek_1976_PhysLett}. 
One coil provided a magnetic field opposite of the guide field that created, in the absence of the other coil,  a zero field region in the center of the coils.  
A second coil created a  magnetic field  
 \begin{eqnarray}
B_\mathrm{f} = \frac{\pi^2\hbar^2}{\mu_\mathrm{n} m_\mathrm{n}\lambda}L^{-1} \label{Eq_Proc_2}
\end{eqnarray}
perpendicular to the neutron polarization direction.
The field $B_\mathrm{f}$  was tuned such that the neutron undergoes half a Larmor precession cycle.
Here $L$ is the active length inside the coils, $m_\mathrm{n}$ is the neutron mass, $\vec{\mu_\mathrm{n}}=\gamma \hbar \vec{\sigma}_\mathrm{n}$ is the neutron magnetic dipole moment, and
$\gamma$ is the gyromagnetic ratio.
When energized the precession coil spin flipper allowed the neutron spin state to be rotated $\pi$ radians with nearly \unit[100]{\%} efficiency. 
\par
Helmholtz coils \unit[56]{cm} in diameter were centered on the target cell and provided a  field of \unit[1.5]{mT}.  
To preserve the neutron  polarization between the supermirror polarizer and the Helmholtz coils a series of permanent magnets provided a magnetic guide field with a minimum of \unit[1]{mT}.
The heat output from the Helmholtz coils, which was not actively cooled, increased the temperature variation for this experiment.
The temperature around the interferometer was controlled with heating tape and calibrated platinum resistance sensors in closed loop, proportional-integral-derivative (PID) operation and typically achieved
a temperature stability of  \textpm\ \unit[5]{mK} \cite{Pushin_2008_PRL}.
For this experiment the interferometer enclosure was stable to only \textpm\ \unit[20]{mK} because of the increased heat caused by the Helmholtz coils.
\par
A neutron interferometer consists of a perfect silicon crystal machined so that there are several parallel crystal blades on a common monolithic  base.
The interferometer used here is shown in Fig.~\ref{Fig_Proc_2}. 
Neutrons entering the interferometer Bragg diffract in the first (splitter)  blade of the interferometer. 
This coherently separates the neutron into two spatially separate paths labeled I and II (see Fig.~\ref{Fig_Proc_1}b). 
Both neutron paths are Bragg diffracted in a second (mirror) blade and interfere in the final (analyzer) blade of the interferometer.   
Conceptually, this is analogous to a Mach-Zehnder interferometer in light optics. 
The two beams exiting the interferometer are historically labeled the O- and H-beams.
Neutrons are detected with near \unit[100]{\%} efficiency using  \unit[25.4]{mm} diameter cylindrical \he\ filled proportional counters.  
\par  
The target was a sealed glass cell containing \he\ gas  (see Fig.~\ref{Fig_Proc_3}) placed in path I of the interferometer.  
A phase flag, which consisted of \unit[2]{mm} thick quartz, was placed in both paths of the interferometer.  
Rotating the phase flag by an angle $\epsilon$ varied the relative phase shift between the two neutron paths and thus modulated the intensity in the O- and H-beam detectors. 
For the O- and H- beam detectors, the intensity can be written for a neutron with spin state $\uparrow(\downarrow)$ as
\begin{eqnarray}
I_0(\epsilon) &=&c_0^{\uparrow(\downarrow)} + c_1^{\uparrow(\downarrow)}\cos\left[\phi_\mathrm{flag}(\epsilon)+ \phi^{\uparrow(\downarrow)}\right] \label{Eq_Proc_3}  \\
I_H(\epsilon) &=&c_3^{\uparrow(\downarrow)} - c_1^{\uparrow(\downarrow)}\cos\left[\phi_\mathrm{flag}(\epsilon)+ \phi^{\uparrow(\downarrow)}\right]  \label{Eq_Proc_4} \\
\mbox{where } I_0& +&I_H =\mbox{constant.} \label{Eq_Proc_5}
\end{eqnarray}
The coefficients $c_i^{\uparrow(\downarrow)}$ for $i=0,1, \ldots, n$ are treated as fit parameters.  
Here $[\phi_\mathrm{flag}(\epsilon)+ \phi^{\uparrow(\downarrow)}]$ is the relative phase difference between the two paths.  
The phase $\phi^{\uparrow(\downarrow)}$ includes both $\phi_\mathrm{sam}$ and any initial relative phase difference between  paths I and II.  
The term $\phi_\mathrm{flag}(\epsilon)=c_2f(\epsilon) $ is the phase shift due to the phase flag where
\begin{eqnarray}
f(\epsilon)=\frac{\sin\left(\theta_B\right)\sin\left(\epsilon\right)}{\cos^2\left(\theta_B\right)-\sin^2\left(\epsilon\right)} \label{Eq_Proc_6}
\end{eqnarray}
is the difference in optical path length for path I and II.   
Here $\theta_B = $ \unit[37.73]{\degree} is the Bragg angle for the interferometer. 
Since \he\ has spin-dependent absorption, $c_i^\uparrow \neq c_i^\downarrow$ for $i=0,1,3$.
\par
The  contrast or fringe visibility $C$ of the interferometer  is the ratio of the amplitude $c_1$ and mean $c_0$ in \eq{Eq_Proc_3}.  
In practice the contrast for a typical neutron interferometer is less than unity due to a host of reasons including small crystal imperfections and temperature gradients. 
Under the best experimental conditions neutron interferometers have demonstrated at most around \unit[90]{\%} contrast.
In this experiment there are two losses of contrast that, although not unique, are of particular interest.
\par
The first case is due the interaction of the neutron as it passes through the glass windows of the target cell. 
As a neutron passes through the glass windows it experiences a large phase shift $\phi_\mathrm{win}$.  
Although this phase shift $\phi_\mathrm{win}$  is spin-independent and is canceled when subtracting the phase measured  in both  neutron spin states, $\phi_\mathrm{win}$  does affect the measured contrast and hence overall  precision of  the experiment.  
This is because the incident neutron beam contains a small  wavelength spread of $\sigma_\lambda/\lambda =$ \unit[1]{\%}.
Neutrons of slightly different wavelengths will experience slightly  different $\phi_\mathrm{win}$ which dephases the detectable  neutron beam after the interferometer.  
This doesn't affect the measured phase  determined  by \eq{Eq_Proc_3} but dephasing does decrease the contrast. 
Assuming a Gaussian spectrum of $\lambda$,  the measured contrast becomes $C=C_{0}\exp{[-(N_\mathrm{win}D_\mathrm{win}b_\mathrm{win}^\prime\sigma_\lambda)^2/2]}$ where $C_0$ is the initial contrast of the interferometer. 
Here $N_\mathrm{win}$, $D_\mathrm{win}$, and $b^\prime_\mathrm{win}$ are the density, thickness, and effective scattering length of the window, respectively. 
This effect is negligible for the \he\ itself because the density of the gas is much lower.
A more  complete description of coherence and subsequent contrast loss can be found in a number of sources including \cite{Clothier_1991_PhysRev,Kaiser_1983_PRL,Klein_1983_PRL,Petrascheck_1988_Physica,Rauch_1996_PhysRev,Pushin_2008_PRL}. 
To minimize the loss of contrast from the cell windows, \unit[8]{mm} of  compensating glass made from 2 target cell windows was placed in path I of the interferometer.
\par
Another mechanism of contrast loss in this experiment is due to the fact that the \he\ sample is a neutron absorber; therefore both $c_0$ and $c_1$ are decreased from what they would  otherwise be in an empty interferometer.
For absorption we have \cite{Sears_1989_Book},
\begin{eqnarray}
c_{0'}&=&\frac{c_0}{2}(1+e^{-\sigma_a N_3 D_3})  \label{Eq_Proc_7}\\
c_{1'}&=& c_1 e^{-\sigma_a N_3 D_3/2}  \label{Eq_Proc_8}\\
C=\frac{c_{1'}}{c_{0'}}&=&C_{0}\sech(\sigma_a N_3D_3/2)  \label{Eq_Proc_9}
\end{eqnarray}
 Absolute contrast during this experiment was a function of the environment,  effectiveness of compensation glass, neutron spin-state and the \he\ polarization  and varied  between \unit[30]{\%} and  \unit[75]{\%}.

\subsection{Glass Target Cells \label{SubSec_Cells}}
\begin{figure}[t]
	\centering
	\includegraphics[width=3.375in]{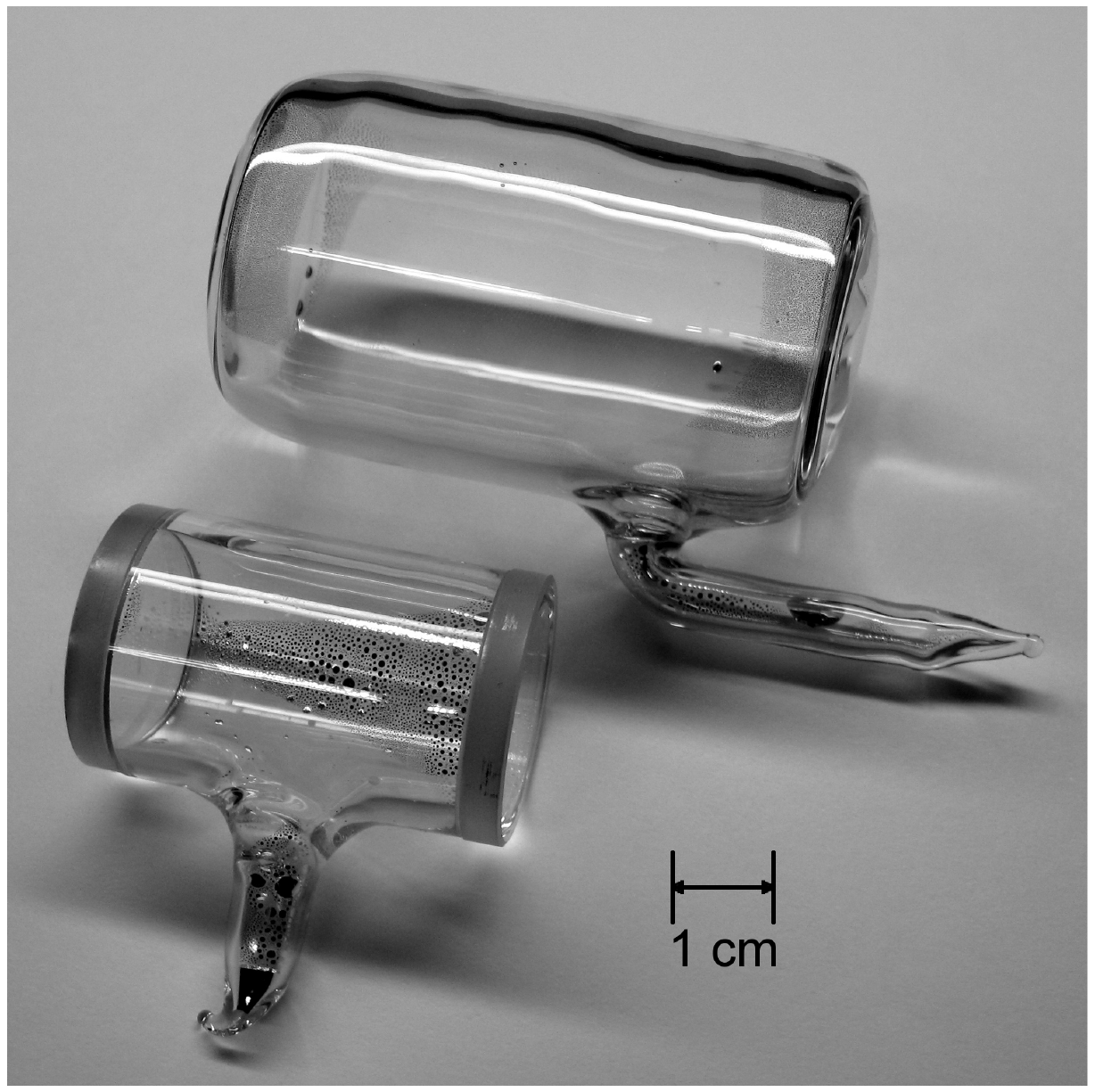}
	\caption{Two \he\ cells. The larger of the two cells, called Skylight, was used in the polarimetry measurements. The smaller flat-window cell called Pistachio was used as a target cell. Rubidium deposits can be seen as dark spots along the walls of the cell.} 	
	\label{Fig_Proc_3}
\end{figure}

The NIST optical shop fabricated four geometrically identical flat-windowed \he\ cells.
Two of these cells, named Cashew and Pistachio  (Pistachio is shown in Fig.~\ref{Fig_Proc_3}) were used in this experiment.   
Each cell was made from  boron-free aluminosilicate  glass \cite{Gentile_2007_JAmerGlassSoc} and consisted of two flat \unit[25]{mm} diameter, \unit[4]{mm} thick windows fused onto a \unit[34]{mm} long cylindrical base.  
Before the cells were sealed they were filled with between \unit[1.7]{bar} and \unit[1.9]{bar} of \he\ gas. 
Nitrogen and rubidium were also added  in order to polarize the \he\ using spin exchange optical pumping (SEOP) \cite{Walker_1997_RMP}.   
Properties of the cells can be found in Table \ref{table1}.  
\par
The environmental constraints at the NIOF required that the cells be polarized at a separate facility.   
In this  facility SEOP was employed to polarize the \he\ gas over a period of \unit[2]{days} to an initial \he\ polarization between  \unit[65]{\%} and \unit[75]{\%}. 
It was also possible at the SEOP facility to monitor and flip the \he\ polarization by nuclear magnetic resonance (NMR) techniques \cite{Bloch_1946_PhysRev}.  
The cells were transported to the NIOF using a portable battery powered solenoid. 
Losses in $^3$He polarization  from transporting the cell between the two facilities were measured to be  \unit[$<2$]{\%}.   
Helmholtz coils placed around the interferometer provided a uniform magnetic field which minimized the loss of \he\ polarization due to magnetic field gradients.   
\begin{table}{\caption{The \he\ cells properties. 
		Quoted spin relaxation times are for the interferometer facility which had  magnetic field gradients. 
		$N_{3}\sigma_{p}D_{3}$ is the opacity of the cell.  
		The pressure at \unit[26]{$\degree$C} was determined assuming that the transmission through the cell windows was 88\,\% \cite{Chupp_2007_NIaMiPRSAASDaAE}. 
    The cell Cashew's relaxation time was \unit[135]{h} in 2008 and  \unit[150]{h} in the 2013 data set.  Pistachio was used only in 2008. 
		\label{table1} }}
		\centering 
		\begin{tabular}[c]{lc@{\hspace{0.075in}}c@{\hspace{0.075in}}cc}\hline\hline 
     		     & Relaxation & $N_{3}\sigma_{p}D_{3}$ & Pressure  &   \\
		Cell Name &  time [h]      & at 2.35 \AA &[bar]&	Function		\\\hline 
		Cashew ('08)    & 135      & 1.1	    & 1.9 & Target   \\
		Cashew ('13)    & 150     & 1.1	    & 1.9 & Target   \\
		Pistachio & 35       & 1.0      & 1.7 & Target  \\
		\hline
		Skylight  & 110     & 3.1	    & 3.1 & Polarimetry   \\
		Haystack  & 80    & 3.0      & 2.94 & Polarimetry    \\
		Whiteface & 35      & 3.6      & 3.50 & Polarimetry  \\\hline\hline
		\end{tabular}		
\end{table}
Cell relaxation times at the interferometer varied per cell with a maximum of \unit[150]{h}.


\subsection{Phase Data \label{SubSec_Phase}}
\begin{figure}[t]
	\centering
	\includegraphics[width=3.375in]{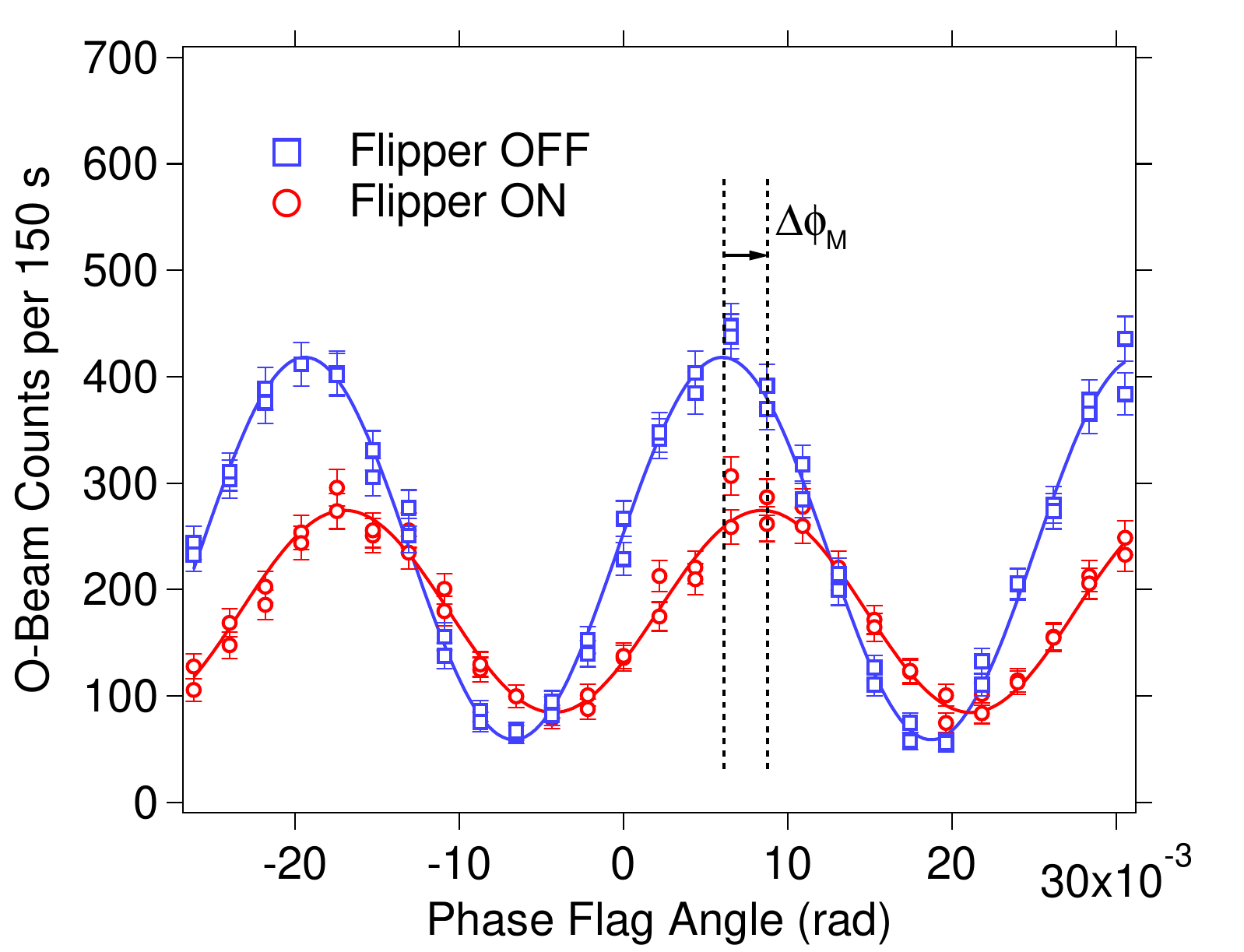}
	\caption{(Color online)  Typical interferograms generated by a off-on-on-off  spin flipper sequence is shown where the spin flipper is off (blue squares) and on (red circles). 
	Each point was counted for \unit[150]{s}. 
	The lower intensity for the ``Flipper ON'' interferogram is due to stronger absorption in that case.  
	The uncertainties shown are purely statistical.  
	Lines are fits of the data using \eq{Eq_Proc_3}. }
	\label{Fig_Interferogram}
\end{figure}

\begin{figure}[t]
	\centering
	\includegraphics[width=3.375in]{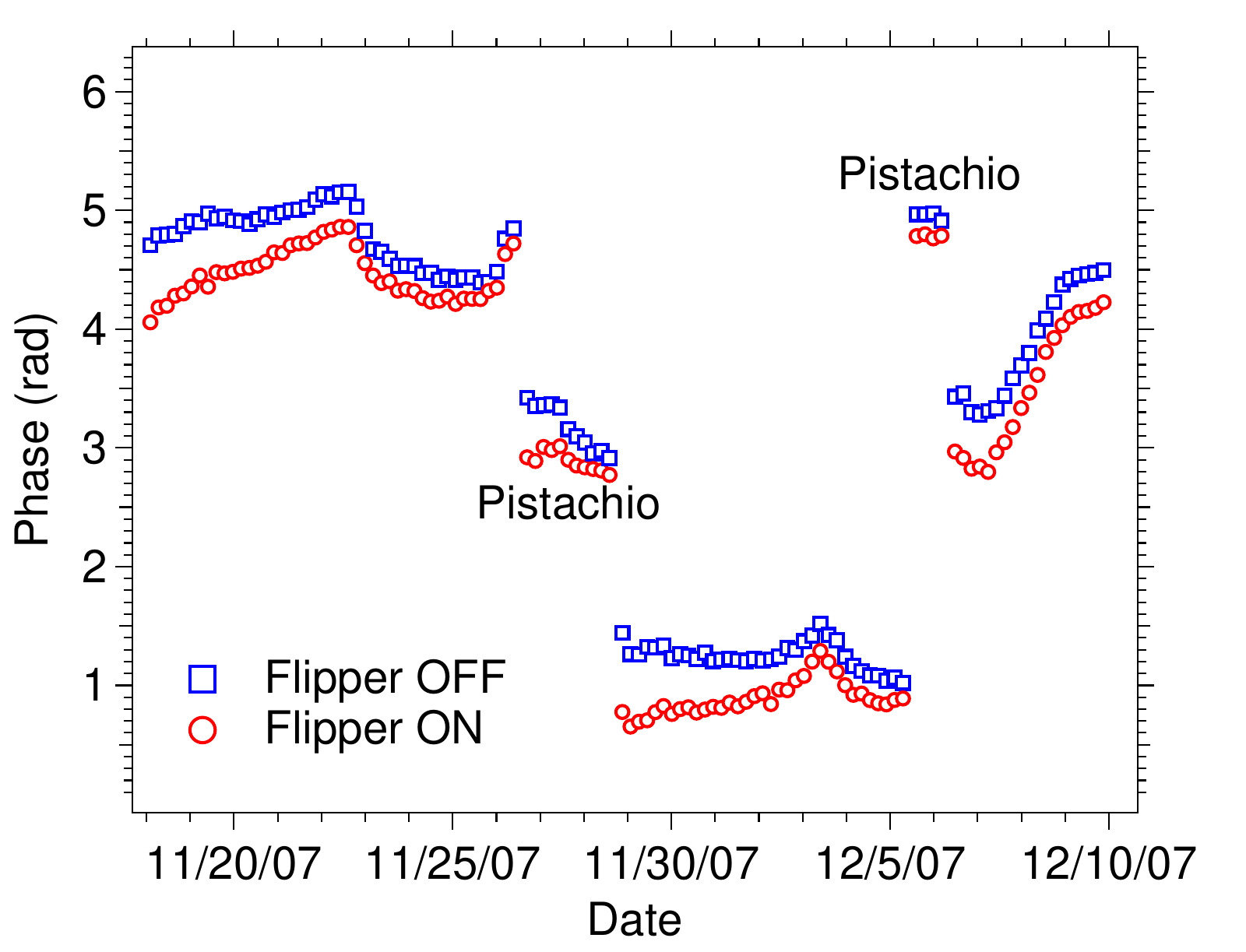}
	\caption{(Color online) The measured phases $\phi_1^{\uparrow}$ (blue squares)  and $\phi_1^{\downarrow}$ (red circles) for a subset of the data that includes 5 cell transfers.  The data taken with the cell Pistachio are marked.  The statistical uncertainties are smaller than the points. }
	\label{Fig_Phase}
\end{figure}

The phase shift caused by the spin-dependent interaction with the target \he\ was measured by rotating the phase flag from an angle of $\epsilon = $ \unit[0]{mrad} to $\epsilon_\mathrm{max}$ and then from  $\epsilon=\epsilon_\mathrm{min}$ to \unit[0]{mrad} in \unit[2.18]{mrad} steps.  
The angles $\epsilon_\mathrm{max}$ and $\epsilon_\mathrm{min}$ varied slightly per run with $\epsilon_\mathrm{max}-\epsilon_\mathrm{min}=$ \unit[56.68]{mrad}.
Each run lasted \unit[4]{h} to \unit[9]{h} with a statistical mode of \unit[4.4]{h}.  
At each angle of the phase flag the spin flipper was operated in a off-on-on-off sequence to reduce the effect of linear drifts. 
Two simultaneous interferograms, one for each precession coil spin flipper state, were constructed from the background subtracted off-on-on-off data. 
A typical pair of interferograms are shown in Fig.~\ref{Fig_Interferogram}. 
Figure \ref{Fig_Phase} shows the measured phases $\phi_1^{\uparrow}$ and $\phi_1^{\downarrow}$ over a span of a month that includes five cell transfers.    

Comparing the two interferograms yields a measured phase shift $\Delta\phi_M$.  
A correction must be applied to $\Delta\phi_M=\phi_1^{\uparrow}-\phi_1^{\downarrow}$ in order to determine $\Delta b^\prime$ using \eq{Eq_Proc_1}. 
This is because the incident beam is an incoherent mixture of both spin-up and spin-down neutrons (the neutron polarization $P_n\neq 1$). 
The measured interferogram is actually a sum of two different interferograms.
\begin{eqnarray}
I_0(\mbox{off})&=&c^{\uparrow}_1\cos\left(\phi_\mathrm{flag} +\phi_1^{\uparrow}\right)\nonumber \\ 
&=&\cos\left(\phi_\mathrm{flag} +\phi_2\right)+\eta_{-}\cos\left(\phi_\mathrm{flag} +\phi_ 3\right) \label{Eq_Proc_10}
\end{eqnarray}
where 
\begin{eqnarray}
\eta_{-} &=&\frac{(1-P_n)}{(1+P_n)}e^{-N_{3}\sigma_{p}D_{3}P_3} \label{Eq_Proc_11}
\end{eqnarray}
is the ratio of the number of minority-spin neutrons to the number of majority-spin neutrons that exit the \he\ target cell. 
In \eq{Eq_Proc_10} the mean intensity has been subtracted but  this does not affect the overall result below.  
When the precession coil spin flipper is energized
\begin{eqnarray}
I_0(\mbox{on}) &=& c^{\downarrow}_1\cos\left(\phi_\mathrm{flag} +\phi_1^{\downarrow}\right)\nonumber\\
&=& \eta_{+}\cos\left(\phi_\mathrm{flag} +\phi_2\right)+\cos\left(\phi_\mathrm{flag} +\phi_ 3\right)\label{Eq_Proc_12}
\end{eqnarray}
where 
\begin{eqnarray}
\eta_{+} &=&\frac{(1-sP_n)}{(1+sP_n)}e^{+N_{3}\sigma_{p}D_{3}P_3} \label{Eq_Proc_13}
\end{eqnarray}
is again the ratio of the number of minority-spin to majority-spin neutrons. 
We can now write the measured phase shift $\Delta\phi_M$ in terms of $\Delta\phi_0=\phi_2-\phi_3$ which is the phase shift if the neutron polarization had been perfect (\unit[100]{\%}) using
\begin{eqnarray}
\Delta\phi_M &=&\arctan{\left(\frac{\displaystyle\sin{\Delta\phi_0}}{\displaystyle\eta_{+}+\cos{\Delta\phi_0}}\right)}\nonumber \\
&\quad &-\arctan{\left(\frac{\displaystyle\eta_{-}\sin{\Delta\phi_0}}{\displaystyle1+\eta_{-}\cos{\Delta\phi_0}}\right).} 
\label{Eq_Proc_14}
\end{eqnarray}
No correction is necessary for the fact that the helium polarization $P_3 \neq 1$  because each individual neutron interacts with multiple \he\ atoms.
\par
We have collected two sets of  \nhe\ phase shift data taken in separate years. 
The first run of this experiment done in 2008 and has previously been reported in Huber \etal \cite{Huber_2009_PRLa,Huber_2009_PRL}.  
A second data set consisting of six months of additional phase measurements was taken in the spring and summer of 2013.

\subsection{Measuring cell relaxation \label{SubSec_Relax}}
\begin{figure}[t]
	\centering
	\includegraphics[width=3.375in]{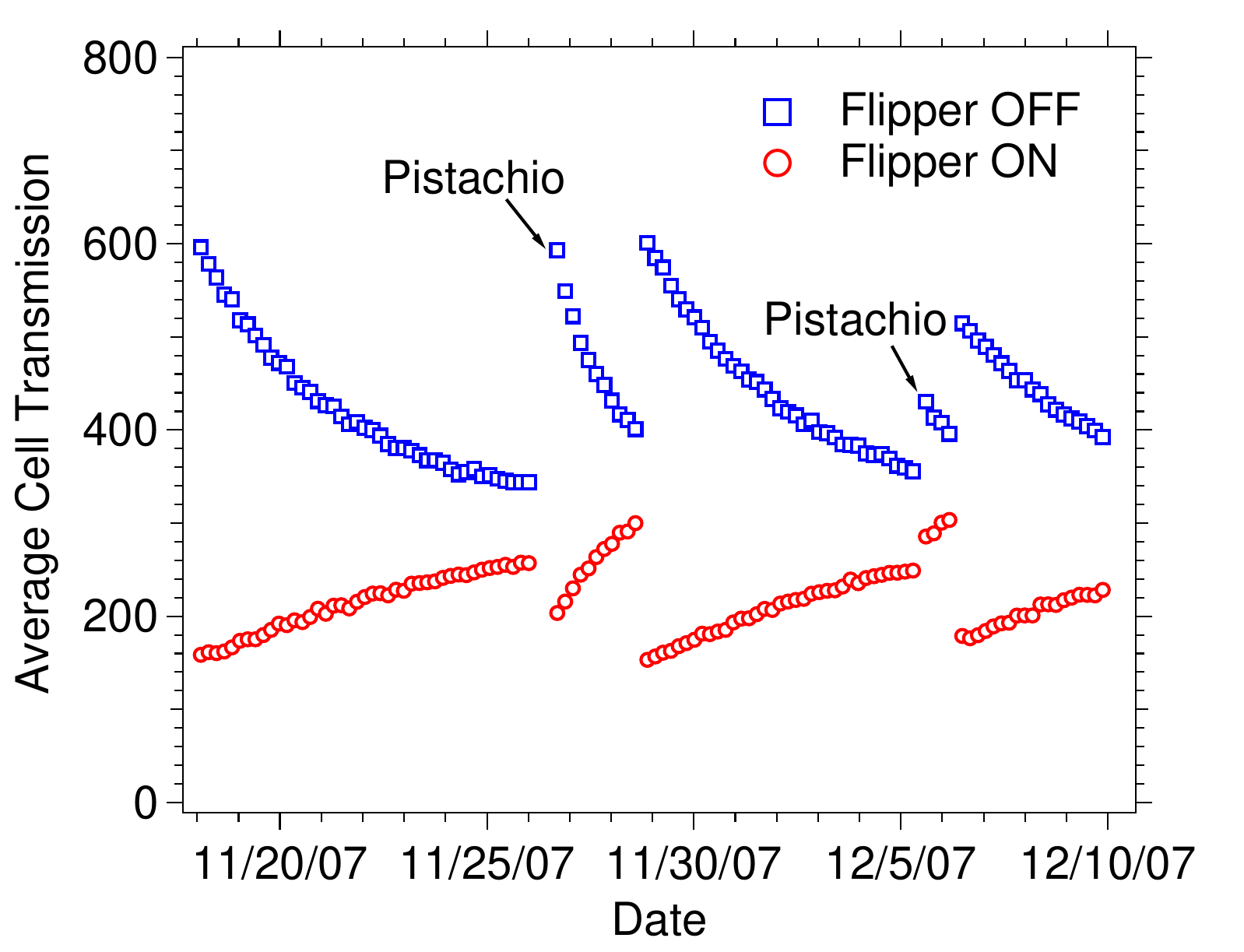}
	\caption{(Color online) The average cell transmissions $I^{\uparrow}$(blue squares) and $I^{\downarrow}$ (red circles) as measured by the C4 detector for a subset of the data that includes 5 cell transfers.  The data taken with the cell Pistachio are marked.  The statistical uncertainties are smaller than the points. }
	\label{Fig_CellTrans}
\end{figure}

Target cell transmission was measured \emph{in situ} during each scan using the C4 detector (see Fig~\ref{Fig_CellTrans}).  
For each run the  asymmetry 
\begin{eqnarray}
A=\frac{I^{\uparrow}-I^{\downarrow}}{I^{\uparrow}+I^{\downarrow}} \label{Eq_Proc_16b}
\end{eqnarray}
 was calculated from the individual off-on spin flipper asymmetries.   
The asymmetry is related to the neutron polarization  $P_n$ and spin flipper efficiency   
\begin{eqnarray}
s =\abs{\frac{P_n(\mbox{on})}{P_n(\mbox{off})}}, 
\end{eqnarray}
where on(off) refers to the state of the precession coil spin flipper \cite{Wildes_1999_RevSciInst}, 
by
\begin{eqnarray}
A=\frac{(1+s)P_nP_A}{2+(1-s)P_nP_A}. \label{Eq_Proc_16}
\end{eqnarray}
The values of $s$ and $P_n$ are known from the polarimetry measurements.  
For each interferogram an averaged asymmetry $\overline{A}$ was calculated. 
The analyzing power $P_A$ of a \he\ cell can be written \cite{Rich_2002_NIM}
\begin{eqnarray}
P_A =\tanh\left(\xi\right) \label{Eq_Proc_17}
\end{eqnarray} 
where
\begin{eqnarray}
\xi=N_{3}\sigma_{p}D_{3}P_3\label{Eq_Proc_18}
\end{eqnarray} 
is the product of the opacity of the cell $N_{3}\sigma_{p}D_{3}$ and \he\ polarization.  
One can use \eqs{Eq_Proc_16}{Eq_Proc_17} to write $\xi$ in terms of the measured asymmetry $\overline{A}$ as
\begin{eqnarray}
\xi=\tanh^{-1}\left(\frac{2\overline{A}}{(s+1)P_n+(s-1)P_{n}\overline{A}}\right). \label{Eq_Proc_19}
\end{eqnarray} 

\section{Phase shift due to a non-uniform magnetic field \label{sec_field}}
When a magnetic field  $B$  in the direction of the neutron polarization is present, the additional magnetic potential $V_\mathrm{M}=-\vec{\mu_{\mathrm n}}\cdot \vec{B}$ experienced by the neutron creates a phase shift 
\begin{eqnarray}
\phi_\mathrm{mag} = \pm\frac{\mu_{\mathrm n} m_{\mathrm n} \lambda D}{2\pi \hbar^2}B  =\pm\kappa DB. 
\label{Eq_Proc_15}
\end{eqnarray}
where $\kappa=$ \unit[$-$545]{mrad/(cm-mT)}.  The distance $D$ is the path length inside the interferometer. 
The $(+)$ and $(-)$ signs correspond to neutron polarization parallel and anti-parallel to the magnetic field, respectively.  
When calculating the phase difference between precession coil spin flipper states off and on the $\pm$ sign in \eq{Eq_Proc_15} reverses,
hence in the absence of polarized \he\ gas this difference is  $2\phi_\mathrm{mag}$ for each path of the interferometer.
Since the length of both interferometer paths are equal, $\phi_\mathrm{mag}$ can only be nonzero if the magnetic field in the two paths are different.
In this case, the phase shift difference will be 
\begin{equation}
2\phi_\mathrm{mag}=2 \kappa (B_1 - B_2) D
\end{equation}
where $B_1$ and $B_2$ are the magnetic field strengths averaged over paths 1 and 2, respectively.
For this interferometer the longer, parallel part of the beam paths, which contained both the cell and compensating glass,  was \unit[6.4]{cm} with  a total path length of \unit[8.6]{cm}.
A magnetic field gradient will be manifested as a non-zero phase shift in the absence
of polarized gas and a non-zero y-intercept for a fit of the variation of phase shift with \he\ polarization.
\par
By direct measurements without the cell inside the interferometer Huber \etal \cite{Huber_2009_PRLa,Huber_2009_PRL} 
determined $2\phi_\mathrm{mag}=$\werb{2}{10}{mrad} hence consistent with zero but with a relatively large uncertainty.
Both the phase shift data versus helium polarization and an estimate of  $2\phi_\mathrm{mag}$ from \he\ relaxation also yielded  $2\phi_\mathrm{mag}$ consistent with zero.
However, after applying the incoherent beam correction of \eq{Eq_Proc_14} a fit of the phase shift data versus helium polarization yielded a nonzero value of \werb{16}{4}{mrad}. 
Furthermore, we found an error in the estimate of  $2\phi_\mathrm{mag}$ from \he\ relaxation in Ref.~\cite{Huber_2009_PRL}.
This led us to directly map the magnetic field, revisit the estimate from \he\ relaxation, and perform a better evaluation of the y-intercept of $\Delta\phi/P_3$ (see Sec.~\ref{Sec_Result}).
In addition to this re-analysis of our 2008 data, we obtained new data in 2013 with a focus on a better evaluation of the intercept.  
In particular we obtained data for both directions of the \he\ polarization and substantially more data at $P_3 = 0$.
\par
\begin{figure}[t]
	\centering
	\includegraphics[width=3.375in]{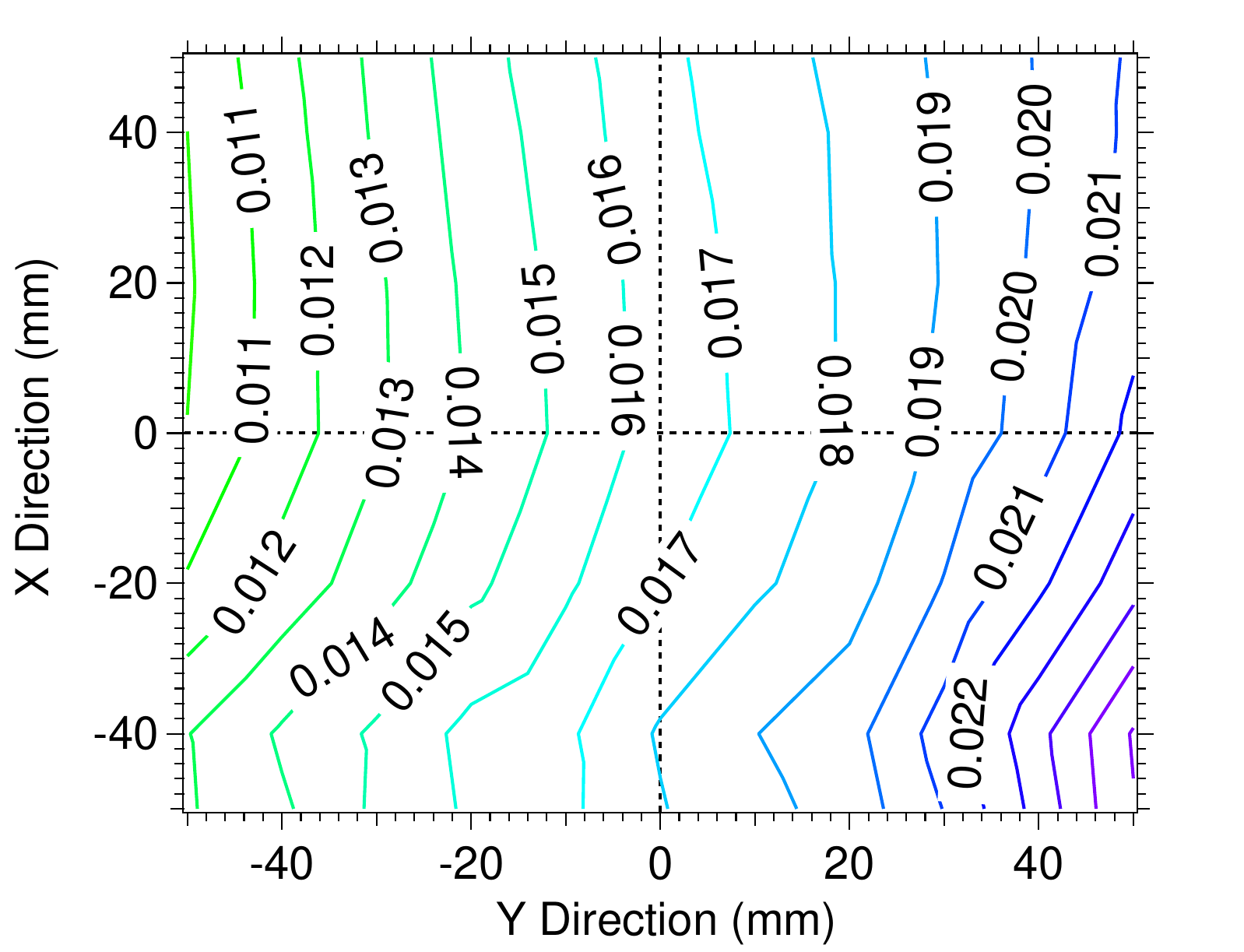}
	\caption{(Color online) The difference between the maximum magnetic field in mT and the magnetic field $B_z$ in the region of the target cell's center (0,0) measured using a fluxgate magnetometer.  Contour lines show a \unit[$7\times10^{-4}$]{cm$^{-1}$} field gradient.  }
	\label{Fig_Mag}
\end{figure}

Figure \ref{Fig_Mag} shows a map of the magnetic field obtained with the Helmholtz coils,
which revealed a fairly linear gradient $(1/B_z) dB_z/dy \approx$ \unit[7$\times10^{-4}$]{cm$^{-1}$}.
The two paths of the interferometer are separated by \unit[2.2]{cm}, hence this gradient
yields $B_1 - B_2 \approx $ \unit[$ 2.3\times10^{-3}$]{mT} and thus $2\phi_\mathrm{mag} \approx $ \unit[16]{mrad}.
Although magnetic parts were avoided near the $^3$He cell, there was a rotation stage with magnetic components 
below the coils.
This stage was necessary so that the interferometer could be aligned to satisfy the Bragg condition.
\par
The observed relaxation time $T_1$ of the $^3$He gas results from contributions from 
dipole-dipole relaxation  \cite{Newbury_1993_PRA}, wall relaxation  \cite{Fitzsimmons_1969_PR},
and magnetic field gradients.  The first two components yield the \textquotedblleft intrinsic\textquotedblright \ relaxation
time of the cell, $T_{\mathrm i}$. The observed relaxation time 
in the interferometer in the presence of a field gradient is 
\begin{eqnarray}
\frac{1}{T_{1}}=\frac{1}{T_{\mathrm i}}+\frac{1}{T_{\mathrm{fg}}}
\label{Eq_Proc_14_15}
\end{eqnarray}
where the gradient contribution is given by\cite{McIver_2009_RoSI}
\begin{eqnarray}
\frac{1}{T_{\mathrm{fg}}}&=& \frac{6700}{p}\left(\frac{\abs{\vec{\nabla}B_x}^2}{B^2}+\frac{\abs{\vec{\nabla}B_y}^2}{B^2}\right)\nonumber\\
&=&\frac{6700\beta^2}{p}  \mathrm{h}^{-1}
\label{Eq_Proc_14_16}
\end{eqnarray}
Here $p$ is the pressure in bar and $B_{x,y}$ are the perpendicular components of the magnetic field where the applied field is in the $z$ direction.  
For the cell Cashew, $T_1 = $ \unit[135]{h} and $T_i = $ \unit[330]{h} which yields $\beta=1.1\times10^{-3}$~cm$^{-1}$.
Whereas $\beta$ includes several components, by using  $\nabla \times \vec B=0$ and by assuming the gradient is dominated by 
the linear gradient observed in the field map, one obtains $(1/B_z) dB_z/dy \approx 1.0\times10^{-3}$~cm$^{-1}$,
$B_1 - B_2 \approx $ \unit[$3.6\times10^{-3}$]{mT} and thus $2\phi_\mathrm{mag}\approx$ \unit[25]{mrad}.
It is likely that this value is an upper limit because several gradient components
contribute to relaxation.  As discussed in Sec.~\ref{Sec_Result}, we obtained y-intercepts of \werb{16}{4}{mrad}
in the 2008 run and \werb{21}{3}{mrad} in the 2013 run, consistent with the estimates
from the field map and $^3$He relaxation.
\par
Between 2008 and 2013  other interferometry experiments were performed at the NIOF.
Changes to the NIOF included different shielding, changes to the polarizer, wavelength changes,  the use of different interferometers and mounting, stages, and a change in the monochromator crystal \cite{M.G.Huber_2014Inpreparation_NI&MiPRSAASDaAE}.
Despite these changes the experimental conditions were reasonably well reproduced. 
Planning for additional phase data at $P_3 = 0$ was  started almost immediately after 2008 as the magnetic field gradient became more of a concern.
The Helmholtz coils and spin flipper were at identical positions  in 2008 and 2013.
Other components of the experiment like the electronics, detectors, the interferometer, cell mounting, and  other system components were the same between the two runs as they were reserved for this work and not otherwise used.
Initial $P_3$  was \unit[10]{\%} higher in 2013 because of advances in helium polarization techniques \cite{Chen_2014_JoAP}. 
Another difference was that in 2008  the polarization direction of  $P_3$ was keep fixed; whereas in 2013 $P_3$ was twice  polarized in the reversed direction.
Lastly, the neutron polarization was \unit[3]{\%} lower (see Section \ref{SubSec_Pol}) in 2013 due to changes in the supermirror alignment. 

\section{Auxiliary Measurements \label{Sec_Aux}}

\subsection{Neutron Wavelength \label{SubSec_Lam}}
Because the skew symmetric interferometer uses a (220) reflection in silicon, it is necessary to eliminate higher order $n = 2,3,\ldots$ reflections from the incident beam.  
Neutrons with wavelengths of $\lambda < $ \unit[2.35]{\AA} are poorly polarized by the supermirror and could potentially affect the phase and polarimetry measurements.  
Neutrons with wavelengths corresponding to $n\geq 3$ are suppressed by the liquid hydrogen cold source.  
However a non-negligible amount (\unit[5]{\%}) of $\lambda/2=$ \unit[1.175]{\AA} neutrons are present in the incident beam.  
To eliminate these neutrons a pyrolytic graphite filter \cite{Shirane_1970_NIM} consisting of nine separate PG crystals of varying thickness (\unit[50]{mm} overall) was placed upstream of the supermirror polarizer.  
Neutrons of wavelength \unit[1.175]{\AA} are preferentially reflected by the (114) plane of the graphite and are absorbed by a surrounding boron shield.  
\par
A measurement of the fraction of $\lambda/2$ neutrons was performed with the interferometer removed and a nearly-perfect silicon analyzer crystal (NPC) placed in the direct beam before the interferometer.  
This crystal analyzer is denoted as ``nearly-perfect'' because it contains a small mosaic spread (a small variation in lattice vector direction throughout the crystal)  of \unit[$3.5\times10^{-4}$]{rad}.
The mosaic of the crystal allows a greater fraction of incident neutrons to satisfy the Bragg condition; thus  more reflect from the crystal and increase the overall reflected intensity. 
The relative intensity of $(I_{\lambda/2})/I_\lambda$  was measured by rotating the ``nearly-perfect'' crystal $\pm\theta_B$ and examining the reflected beams. 
In addition to the NPC, a disk chopper made from a rotating, neutron absorbing cadmium disk with a small slit was used as well.
The disk chopper allows time-of-flight analysis of the neutron spectrum.   
Both techniques  placed an upper limit of $(I_{\lambda/2})/I_\lambda <0.1\%$   that was determined by comparing the relative intensities with and without the filter in place.  
This ratio is mainly limited by the accuracy in determining the small background signal. 
The presence of \unit[1.175]{\AA} neutrons at this level had a negligible effect on $\Delta\phi_M$ and polarimetry measurements. 

\subsection{Polarimetry \label{SubSec_Pol}}
Several neutron polarimetry measurements were made throughout the experiment to verify that the neutron polarization was stable over the duration of the experiment. 
Each polarimetry measurement took place during pauses in collection of the phase data.  
Common techniques to measure neutron polarizations and spin flipper efficiencies along with their difficulties are described in Ref.~\cite{Yerozolimsky_1999_NIM}. 
In this experiment the neutron polarization $P_n$ was measured with  \he\ cells  and by using two analysis methods, which we refer to as the asymmetry and normalized transmission methods. 
Polarimetry cells are physically larger than target cells and one of them is shown in Fig.~\ref{Fig_Proc_3}.
Properties of the cells are listed in Table \ref{table1}.
\par
A \he\ analyzer had two advantages over crystal or supermirror analyzers.  
First, the analyzing power $P_A$ of the cell was determined from unpolarized neutron transmission measurements.  
Second, we could flip the cell's polarization by $\pi$ radians using nuclear magnetic resonance (NMR) at the SEOP facility. 
This NMR induced flip eliminated the need for a second spin flipper to uniquely determine $P_n$, $P_A$, and spin flipper efficiency.
These cells had three times the opacity of the target cells and thus provided high analyzing powers that were relatively insensitive to variations in \he\ polarization. 
\par
The setup for both methods was the same.  
Low neutron fluence rates in the H-beam prevented any practical polarization analysis behind the interferometer. 
Instead, the interferometer was removed from its cradle and replaced with one of the analyzing cells.  
Because the neutron polarization produced by the supermirror polarizer should depend very weakly on wavelength and the beam spectrum was sufficiently narrow ($\sigma_\lambda/\lambda=$ \unit[1]{\%}), the difference between the measured $P_n$ of the direct beam and the neutron polarization of paths I and II of the interferometer is believed to be negligible.  
The neutron transmission through the cell was measured using a \he\ detector located directly behind the analyzer.  
\par
For both methods the analyzing power of the cell was determined by the transmission of unpolarized neutrons.  
The analyzing power of a \he\ cell is given by \eq{Eq_Proc_17}.
For the polarimetry cells the range of initial $P_A$ was between 86 \% and 99 \% but was typically around 97 \% depending on the cell and its initial \he\ polarization.
\eq{Eq_Proc_17} can be rewritten as a ratio of two unpolarized neutron transmission measurements as
\begin{eqnarray}
P_A=\sqrt{1-\bigg(\frac{T_\mathrm{un}}{T_\mathrm{pol}}\bigg)^2} \label{Eq_Aux_1}
\end{eqnarray}
where $T_\mathrm{pol}$($T_\mathrm{un}$)  is the transmission of unpolarized neutrons through a polarized (unpolarized) \he\ cell. 
These transmissions are discussed later in Sec.~\ref{SubSubSec_NT}.
Unpolarized neutrons were obtained by translating the supermirror out of the beam using an encoded linear stage. 
The position of the supermirror was reproducible to within \unit[1]{\micro m}.  
To measure $T_{\mathrm{un}}$ the analyzer cell was depolarized by temporarily connecting the Helmholtz coils to an alternating current voltage supply.  

\subsubsection{Asymmetry Method \label{SubSubSec_Assy}}	
\begin{figure}[t]
	\centering
		\includegraphics[width=3.375in]{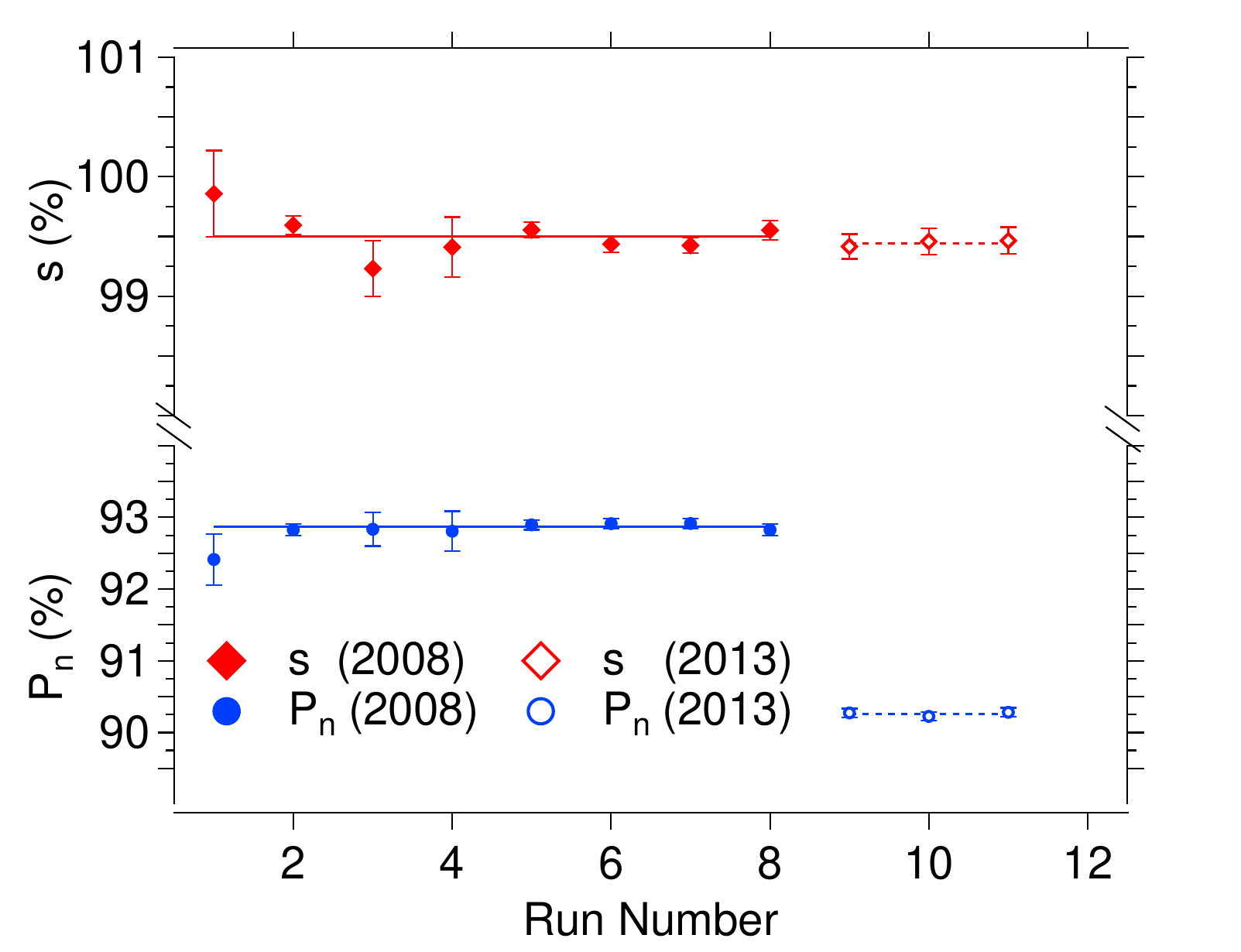}
	\caption{(Color online)  The neutron polarization (circles) and spin flipper efficiency (diamonds)  measured  in  2008 (solid) and 2013 (open)  determined using the asymmetry method. The uncertainties shown are purely statistical.  
Fits of the data are shown as solid lines for 2008 and dotted lines for 2013.  Points with larger uncertainties were taken when the cell had lower polarization.}
	\label{Fig_Aux_1}
\end{figure}

The asymmetry method used the difference in count rates for the two neutron spin states, $I^{\uparrow}$ and $I^{\downarrow}$, to determine the neutron polarization and spin flipper efficiency. 
Here $I^{\uparrow(\downarrow)}$ is the intensity when the neutron and \he\ polarization are aligned parallel (anti-parallel).  
The asymmetry $A$ is related to the neutron polarization and spin flipper efficiency by
\begin{eqnarray}
A&=&\frac{I^{\uparrow}-I^{\downarrow}}{I^{\uparrow}+I^{\downarrow}}= \frac{(1+s)P_nP_A}{2+(1-s)P_nP_A}. \label{Eq_Aux_2}
\end{eqnarray}
To uniquely determine $P_n$ and $s$ using this method it is necessary to have two separate asymmetries $A$ and $A^\ast$ where one reverses the direction of the \he\ polarization.
Similar to \eq{Eq_Aux_2} we have
 \begin{eqnarray}
A^\ast &=& \frac{(1+s)P_nP_{A^\ast}}{2-(1-s)P_nP_{A^\ast}} \label{Eq_Aux_2b}
\end{eqnarray}
The analyzing  powers $P_{A}$ and $ P_{A^\ast}$ were not the same because of a few percent loss in $P_3$ caused by performing an NMR induced spin flip and transporting the cell to and from the SEOP facility. 
Using \eqs{Eq_Aux_2}{Eq_Aux_2b} the spin flipper efficiency is
\begin{eqnarray}
s = \displaystyle \frac{A\left[1+ A^\ast \right]P_{A^\ast }+ P_{A} \left[A-1 \right]A^\ast } {A\left[A^\ast-1 \right]P_{A^\ast }+ P_{A} \left[1+ A \right]A^\ast}.
	\label{Eq_Aux_3}
\end{eqnarray} 
With knowledge of $s$, $P_n$ can be determined via
\begin{eqnarray}
P_n&=&\frac{2A}{P_{A}\left[ (s-1)A + (1+s) \right]}\nonumber\\ 
&=& \frac{2A^\ast}{P_{A^\ast }\left[ (1-s)A^\ast + (1+s) \right]}.\label{Eq_Aux_4}
\end{eqnarray}
Measurements of $P_n$ and $s$ that were obtained using the asymmetry method are shown in Fig.~\ref{Fig_Aux_1}. 
\par
The intensities $I^{\uparrow}$ and $I^{\downarrow}$ for both this method and normalized transmission method discussed in Section \ref{SubSubSec_NT} were performed symmetrically around measurements of $T_\mathrm{pol}$ and hence $P_A$ and $P_{A^\ast }$.
Specifically, $I^{\uparrow}$ (and likewise $I^{\downarrow}$) was measured twice, once before and once after a measurement of $T_\mathrm{pol}$.  
The averages of $I^{\uparrow}$  and  $I^{\downarrow}$  were then used in \eqs{Eq_Aux_2}{Eq_Aux_2b}.
This was done to compensate  for the decay of $P_3$ while the measurements  were being performed to a  level  where no correction for $P_3$ decay was needed.  
\subsubsection{Normalized Transmission Method \label{SubSubSec_NT}}	
When $P_n = 1$ the transmission of neutrons through a polarized \he\ cell is
\begin{eqnarray}
T_\mathrm{off(on)} &=& T_\mathrm{g}\exp{\left(-N_{3}\sigma_aD_{3}\right)} \nonumber \\
&=& T_\mathrm{g}\exp{\left(-N_{3}\sigma_\mathrm{un}D_{3}\right)}\exp{\left({\pm \xi}\right)}, \label{Eq_Aux_5}
\end{eqnarray}
where \eqs{Eq_Theo_15}{Eq_Proc_18} have been used to relate the absorption cross section $\sigma_a$ to $\xi$.  
The signs $(+)$ and $(-)$ is for `off' and `on' states of the precession coil spin flipper, respectively.
Here we have taken that the initial neutron polarization and \he\ polarization is in the same direction.  
$T_\mathrm{g}$ is the transmission of neutrons through the cell windows. 
\par
For a neutron beam  with $P_n\leq 1$ 
the transmission $T_\mathrm{off}$ becomes 
\begin{eqnarray}
T_\mathrm{off} &=&\left(\frac{T_\mathrm{g}}{2}\right)e^{-N_{3}\sigma_\mathrm{un}D_{3}} \nonumber \\
&\times& \bigg[\left(1+P_n\right)e^{\xi}+\left(1-P_n\right)e^{-\xi}  \bigg]. \label{Eq_Aux_6}
\end{eqnarray}
Eq.~(\ref{Eq_Aux_6}) can  be expressed  more compactly as
\begin{eqnarray}
T_\mathrm{off} &=&T_\mathrm{g}e^{-N_{3}\sigma_\mathrm{un}D_{3}}\left[\cosh\left({\xi}\right)+P_n\sinh\left({\xi}\right) \right].\label{Eq_Aux_7}
\end{eqnarray}
\par
The transmission of unpolarized neutrons through a polarized \he\ cell is given by
\begin{eqnarray}
T_\mathrm{pol}=T_\mathrm{un}\cosh\left({\xi}\right), \label{Eq_Aux_8c}
\end{eqnarray}
where 
\begin{eqnarray}
T_\mathrm{un}=T_\mathrm{g}e^{-N_{3}\sigma_\mathrm{un}D_{3}} \label{Eq_Aux_8b}
\end{eqnarray}
is the transmission of unpolarized neutrons through an unpolarized \he\ cell.
Dividing \eq{Eq_Aux_7} by  \eq{Eq_Aux_8c}  yields
\begin{eqnarray}
\frac{T_\mathrm{off}}{T_\mathrm{pol}}=1+ P_n \tanh\left(\xi\right). \label{Eq_Aux_9}
\end{eqnarray}
It follows from Eqs.~(\ref{Eq_Proc_17}), (\ref{Eq_Aux_1}), and (\ref{Eq_Aux_9}) that 
\begin{eqnarray}
P_n=\frac{\frac{\displaystyle T_\mathrm{off}}{\displaystyle T_\mathrm{pol}}-1}{\sqrt{\displaystyle{1-\bigg(\frac{T_\mathrm{un}}{T_\mathrm{pol}}\bigg)^2}}}. \label{Eq_Aux_10}
\end{eqnarray}
When one energizes the spin flipper one has the anti-parallel case where,
\begin{align}
sP_n =\frac{\displaystyle 1-\frac{T_{\mathrm{on}}}{T_\mathrm{un}}}{\displaystyle \sqrt{1-\bigg(\frac{\displaystyle T_\mathrm{un}}{\displaystyle T_\mathrm{pol}}\bigg)^2}}.\label{Eq_Aux_11}
\end{align}
\begin{figure}[t]
	\centering
\includegraphics[width=3.375in]{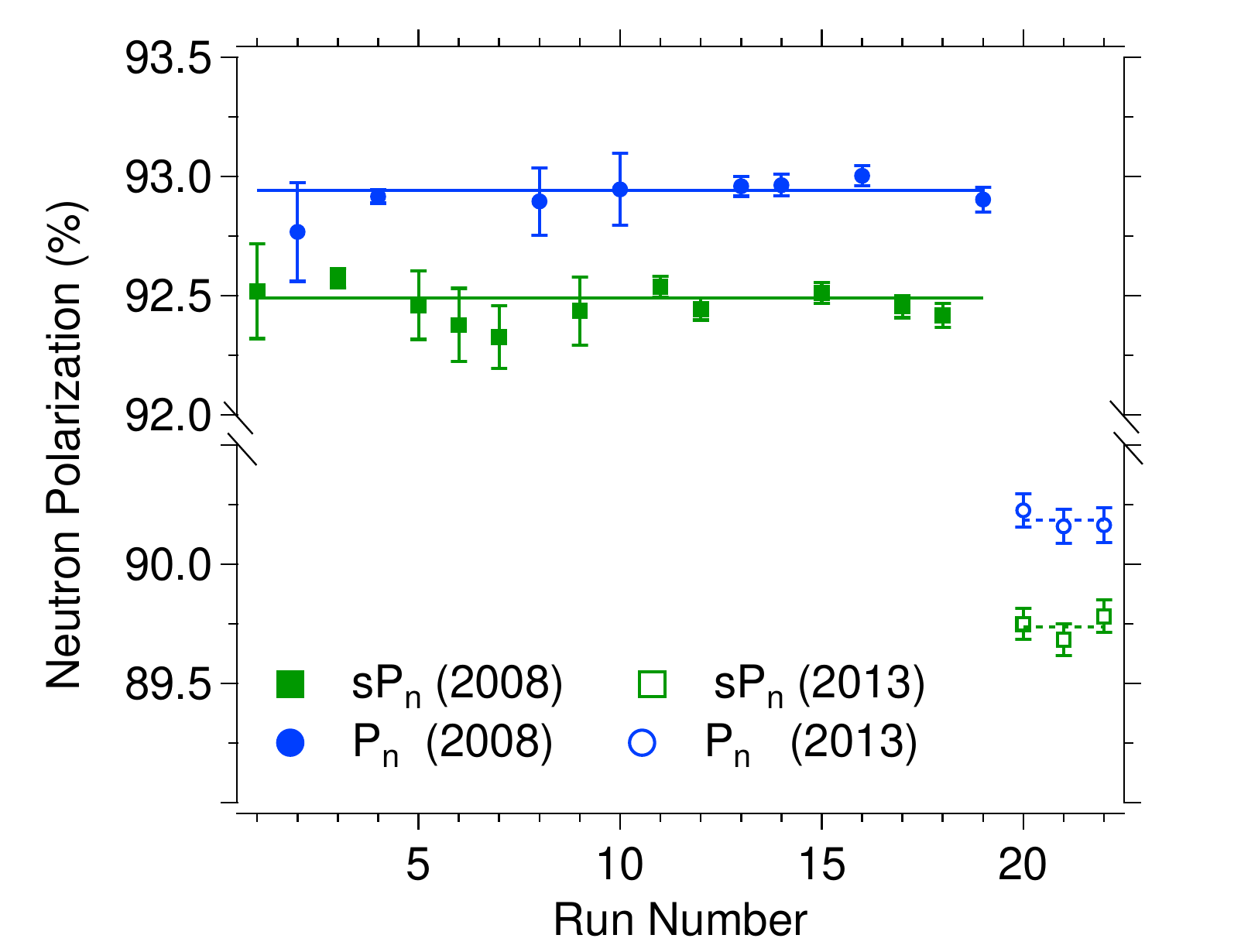}
	\caption{(Color online)  The neutron polarization (circles) and neutron polarization when energizing the precession coil spin flipper (squares)  measured  in  2008 (solid) and 2013 (open)  determined using the normalized transmission  method. The uncertainties shown are purely statistical.  
	All points shown were taken in the anti-parallel state of \nhe\ (see text). 
Fits of the data are shown as solid lines for 2008 and dotted lines for 2013.
Points with larger uncertainties were taken when the cell had lower polarization.}
	\label{Fig_Aux_2}
\end{figure}
\par 
It should be noted that the uncertainty associated with determining the polarimetry from \eq{Eq_Aux_10} is different than that for \eq{Eq_Aux_11}.
Propagating the uncertainty $\sigma$ of the polarization product $Z_\mathrm{off}=P_A P_n$  and $Z_\mathrm{on}=P_A sP_n$ we find
\begin{eqnarray}
\sigma^2_{Z_\mathrm{off}}=\left(\frac{1}{T_\mathrm{un}}\right)^2\sigma^2_{T_\mathrm{off}}+\left(\frac{T_\mathrm{off}}{T_\mathrm{un}^{^2}}\right)^2\sigma^2_{T_\mathrm{un}} \label{Eq_Aux_12}\\
\sigma^2_{Z_\mathrm{on}}=\left(\frac{1}{T_\mathrm{un}}\right)^2\sigma^2_{T_\mathrm{on}}+\left(\frac{T_\mathrm{on}}{T_\mathrm{un}^{^2}}\right)^2\sigma^2_{T_\mathrm{un}}. \label{Eq_Aux_13}
\end{eqnarray}\label{eq31}For this experiment $Z_\mathrm{off}\approx Z_\mathrm{on} \approx 0.9$ and $T_\mathrm{on}\approx 0.05 T_\mathrm{off}$. 
Using \eqs{Eq_Aux_12}{Eq_Aux_13} we find that
\begin{eqnarray}
\sigma_{Z_\mathrm{on}}\approx 0.05\sigma_{Z_\mathrm{off}}.  \label{Eq_Aux_14}
\end{eqnarray}
So despite the relative uncertainties of $T_\mathrm{off}$ and $ T_\mathrm{on}$ being comparable the overall uncertainty in determining $P_n$ versus $sP_n$ differs by a factor of twenty.  The contribution to the overall uncertainty of $P_n$ from the uncertainty in $P_A$ is small because of higher statistics without the supermirror in the neutron beam.  
By reversing the \he\ spin using NMR we can invert \eq{Eq_Aux_14} so that
\begin{eqnarray}
\sigma_{Z_\mathrm{off}}\approx 0.05\sigma_{Z_\mathrm{on}}  \label{Eq_Aux_15}
\end{eqnarray}
when the \he\ polarization has been flipped.
It follows that using the anti-parallel state to determine $P_n$ or $sP_n$ is advantageous despite a much smaller $T_\mathrm{off(on)}$ being measured because the overall uncertainty is better.   
Thus we have chosen to use only the anti-parallel measurements for the normalized transmission method (shown in Figure \ref{Fig_Aux_2}).
\subsubsection{Polarimetry Result \label{SubSubSec_PolRes}}
Both the asymmetry and normalized transmission methods yield neutron polarizations and spin flipper efficiencies to less than \unit[0.1]{\%} relative standard uncertainty. 
These results  are shown in Table \ref{table2}.
For the 2008 data set there was a $2\sigma$ disagreement in measured neutron polarization between the two methods.   
To handle this discrepancy, the uncertainties for the 2008 polarimetry results are determined by adding the largest uncertainty of the two methods in quadrature with the difference between the methods, for example $\sigma_{P_n}=\sqrt{\sigma^2_\mathrm{largest}+(\Delta P_n)^2}=\sqrt{(0.0033)^2+(0.92874-0.92941)^2}$.  
In 2013 the polarimetry  (see Figs.~\ref{Fig_Aux_1} and \ref{Fig_Aux_2}) data were more consistent and this expansion of their uncertainty was not done.  
An equal weighted average of the asymmetry and normalized transmission methods yields 
\begin{align}
P_n & =   0.92908\pm 0.00075   \quad \text{in 2008}\\
&= 		0.90227\pm 0.00055 \quad \nonumber\text{in 2013}
\end{align}
and 
\begin{align}
s & =  0.99510\pm 0.00034  \quad \text{in 2008}\\
	&= 0.99475\pm 0.00089  \quad \nonumber\text{in 2013}
\end{align}
The differences between the 2008 and 2013 neutron polarization is believed to be due to non-reproducible changes in the angular separation between the two mirror surfaces of the supermirror polarizer that is often varied between experiments.  
In both 2008 and 2013 the neutron precession coil spin flipper was the same device,  was located in the same place,  and showed much better agreement.  
\begin{table}{\caption{Results of the polarimetry for the various methods used.   $\sigma_R$ is the relative standard uncertainty. }\label{table2}}
		\centering  
		\begin{tabular}[c]{l@{\hspace{0.05in}}c|@{\hspace{0.05in}}c@{\hspace{0.05in}}c@{\hspace{0.05in}}|c@{\hspace{0.05in}}r}
		\\
		\hline\hline
		                                       &                         & \multicolumn{2}{c}{2008} &\multicolumn{2}{c}{2013}     \\
    \hline
    Var.                       & Method           & Value ($\sigma$) &  $\sigma_R$, [\%] & Value ($\sigma$) &  $\sigma_R$, [\%]  \\
	  \hline
		\multirow{3}{*}{$P_n$} & 	Asy.               & 0.92874(33)        & 0.04   & 0.90260(36)        & 0.04                                                   \\
                                            & 	N.T.               & 0.92941(17)        & 0.02   & 0.90184(41)        & 0.05                                                \\
		                                        & 	Asy. + N.T.    & 0.92908(75)        & 0.09   & 0.90227(55)        & 0.06                                                       \\	
																						\hline
	  \multirow{3}{*}{$s$}     & 	Asy.               & 0.99502(31)        & 0.03   & 0.99444(63)        & 0.06                              \\
	                                          &   N.T.               & 0.99516(23)        & 0.02   & 0.99506(63)        & 0.06                                      \\
                                            & 	Asy. + N.T.    & 0.99510(34)        & 0.03   & 0.99475(89)        & 0.09                                      \\	
 	\hline\hline
		\end{tabular}		
\end{table}

\section{Systematic Effects \label{SubSec_Syst}}

\subsection{Absorption Cross Section \label{Sec_Abs}}
The quantity $\xi$ is a function of $\lambda$ and can be written as
\begin{eqnarray}
\xi=N_{3}\sigma_{p} D_{3}P_3=N_{3}\left[\frac{1}{4}\left(\sigma_0-\sigma_1 \right)\right]\frac{\lambda}{\lambda_\mathrm{th}}D_{3}P_3, \label{Eq_Syst_1}
\end{eqnarray}
where $\lambda_\mathrm{th}= $ \unit[1.798]{\AA} is the reference thermal  neutron wavelength.
To extract $N_{3}\lambda D_{3}P_3$ from $\xi$, one needs the singlet and triplet absorption cross sections $\sigma_0$ and $\sigma_1$.  
The experimental value of $\sigma_\mathrm{un}\approx \sigma_0/4$ is well known from transmission measurements as \werb{5333}{7}{b}  at $\lambda_\mathrm{th}$ \cite{Mughabghab_2006_Book}. 
However, the triplet absorption cross section is poorly known experimentally. 
Passell and Schermer \cite{Passell_1966_PhysRev} measured neutron transmission through \he\ and determined the ratio of singlet to total absorption cross section to be $g_0\sigma_0/\sigma_\mathrm{un} = $ \unit[(1.010$\pm$ 0.032)].
An indirect measurement of the same quantity  was made by Borzakov \etal \cite{Borzakov_1982_SovJNuclPhys} 
where they determined  $g_0\sigma_0/\sigma_\mathrm{un} = $ \werb{0.998}{0.010}{} by examining deviations from a purely ``$1/v$'' absorption law for neutron energies up to \unit[150]{keV}.  
Both of these experiments support $\sigma_1 \approx 0$ but only at the 1 \% level.  
Due to the lack of precision measurements of $\sigma_1$, we used a theoretical prediction of the imaginary part of the scattering length to estimate $\sigma_1$.
\par
Calculations performed by Hofmann and Hale \cite{Hofmann_2003_PhysRev, Hofmann_2008_PhysRev} of the imaginary free scattering length $a_1^{\prime\prime}$ using R-matrix and AV18+3N interactions give a range of values $a_1^{\prime\prime}$ of between \unit[0.0012]{fm} and \unit[0.0051]{fm}.  
However as noted in the same paper, AV18+3N models under-predict the experimentally measured $a_0^{\prime\prime}$ by up to \unit[30]{\%}.  
To be conservative we used $a_1^{\prime\prime} = $ \werb{0.005}{0.005\,}{fm}.  
This allowed for the possibility that theoretical calculations are low by as much as a factor of two.  
With $a_1^{\prime\prime}$ and the measured thermal absorption cross section for unpolarized \he\ we have
\begin{eqnarray}
\sigma_0-\sigma_1 =\mbox{\werb{21236}{100}{b}}. \label{Eq_Syst_2}
\end{eqnarray}
With \eqs{Eq_Proc_19}{Eq_Syst_2} one may extract $N_{3}\lambda D_{3}P_3$ from the asymmetry measurements of $\xi$.

\subsection{Polarimetry Effects \label{SubSec_PolSyst}}
The effect of  uncertainties in $P_n$ and $s$ on calculating $\Delta\phi_0$ is complicated by the fact that they affect both $\eta_{-}$ and $\eta_{+}$ directly and also indirectly through $\xi$.  
To determine the systematic uncertainty in $\Delta\phi_0$ contributed by the uncertainties $\sigma_{P_n}$ and $\sigma_s$ we studied a simulated set of $[\Delta\phi_M]_\mathrm{sim}$ and $[\xi]_\mathrm{sim}$ data.
This simulated data was generated using a fixed value $\Delta b^\prime = $\unit[$-$5.400 ]{fm} and a randomly distributed set of $[\xi]_\mathrm{sim}$ to generate a $[\Delta\phi_M]_\mathrm{sim}$.
$\Delta\phi_0$ was then calculated using the simulated $[\Delta\phi_M]_\mathrm{sim}$ and $[\xi]_\mathrm{sim}$ while varying $P_n$ and $s$  by their respective uncertainties. 
The variance in $\Delta\phi_0$, and hence  $\Delta b^\prime $, resulting from the uncertainties $\sigma_{P_n}$ and $\sigma_s$ was taken as the systematic uncertainty due to the polarimetry measurements.

\section{Results \label{Sec_Result}}
\begin{figure*}[t!]
    \centering
    \begin{subfigure}
		 \centering
        \includegraphics[width=3.375in]{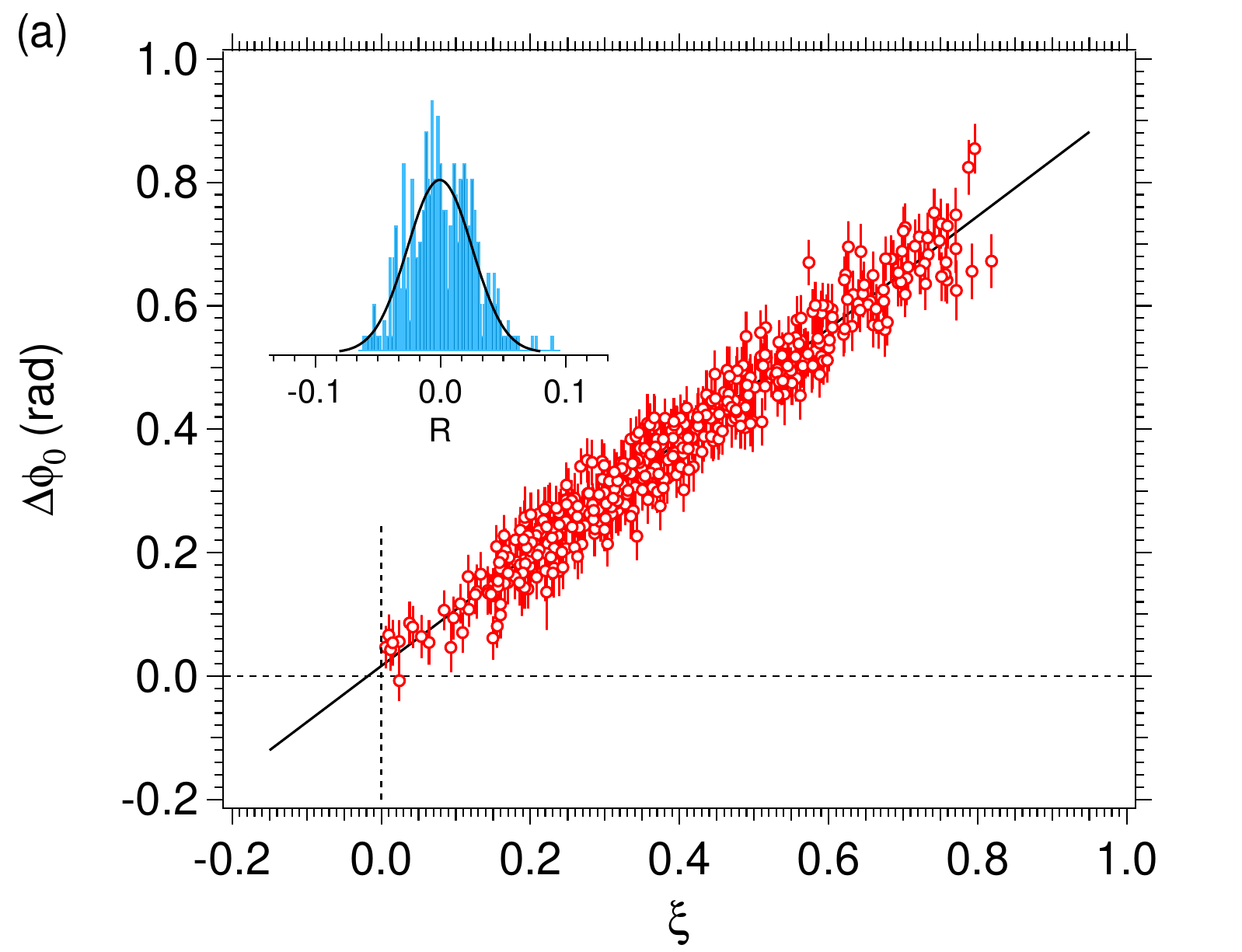} 
    \end{subfigure}%
		~
    \begin{subfigure}
		 \centering
        \includegraphics[width=3.375in]{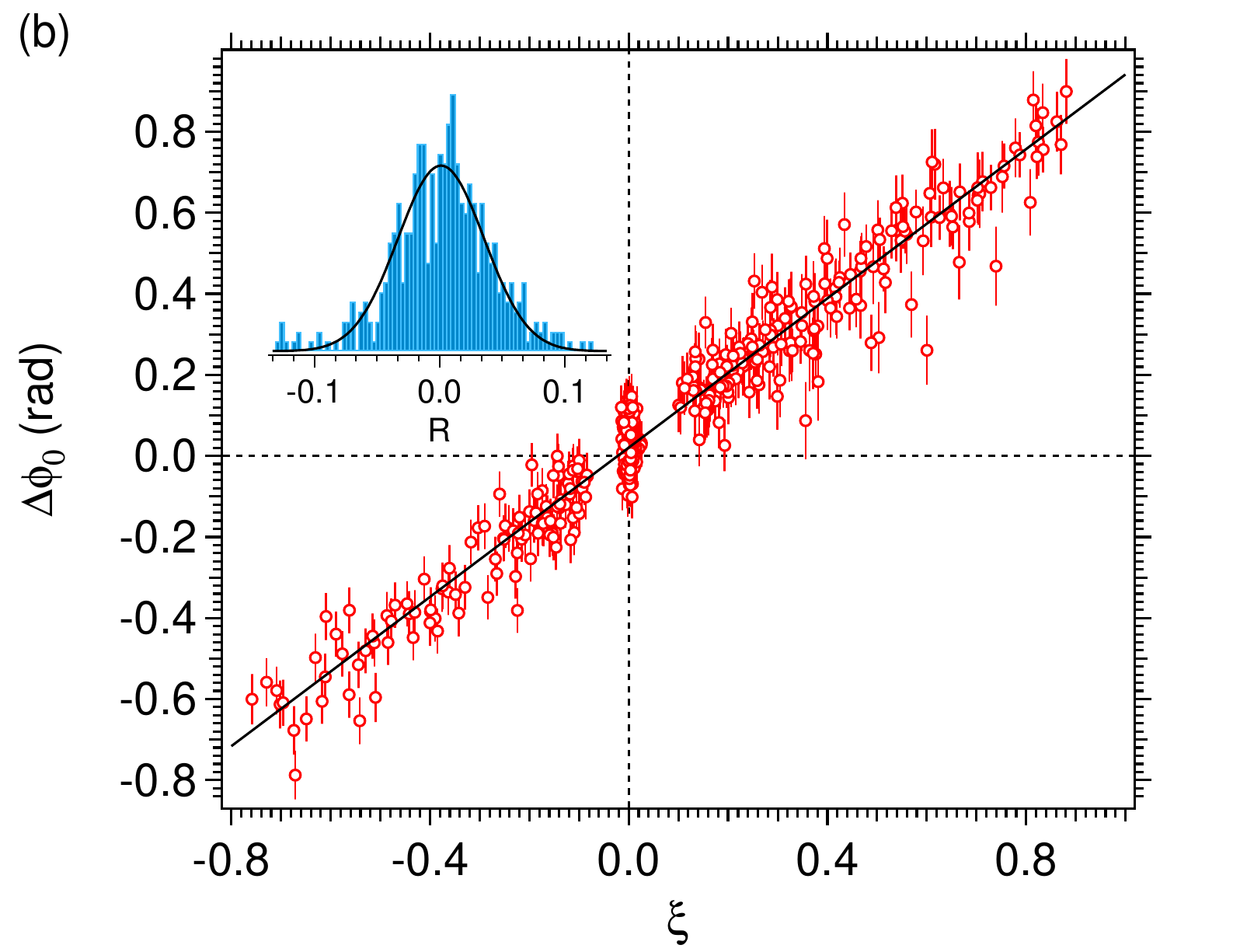}
    \end{subfigure}
    \caption{(Color online) $\Delta\phi_0 $ vs $\xi $ values for (a) 2008 and (b) 2013.  
  The solid line is a weighted average with a $\chi^2/d.o.f.=1$ (see text).  Insets: Histograms of the residual distribution with Gaussian fits (solid lines). The residual is defined by $R=y_i-y$  where $y$ is the fit function and $y_i$ is the $i^{th}$ data point. (a) The Gaussian fit is centered at \unit[$-0.001$]{rad}  with a full width half maximum (FWHM) of \unit[ 0.091]{rad}. (b)  The Gaussian fit is centered at  \unit[$+0.002$]{rad} with  FWHM of \unit[ 0.117]{rad}.\label{Fig_Result_2}}
\end{figure*}

Figure \ref{Fig_Result_2}a shows the measured $\Delta\phi_0$ versus $\xi$ for the 2008 data set which  was collected  over several reactor cycles for a total of 12 weeks. 
From \eq{Eq_Proc_1} the value of $\Delta b^\prime$ can be determined by the slope of $\Delta\phi_0$ in Figure \ref{Fig_Result_2}a.
There are two significant  changes in  determining $\Delta b^\prime$ from what was done previously in Huber \etal \cite{Huber_2009_PRLa,Huber_2009_PRL}. 
The first and most significant is that in Refs.~\cite{Huber_2009_PRLa,Huber_2009_PRL}  the slope of  $\Delta\phi_0$ versus $\xi$ was determined  using a one parameter fit.
This  fixed the y-intercept of the fit to be precisely zero corresponding to $\Delta\phi_0 = 0 $ at $P_3 = 0 $. 
In the presence of a magnetic field gradient, this approach is no longer valid.
Instead we now perform a two parameter fit of $\Delta\phi_0$ versus $\xi$.
The fitted y-intercept of the data shown in Fig.~\ref{Fig_Result_2}a yields  $2\phi_\mathrm{mag} =$ \werb{16}{4}{mrad}.
\par
The other change we have made has been in the manner in which we cut individual data points. 
 In Refs.~\cite{Huber_2009_PRLa,Huber_2009_PRL} we cut the data based on the reduced chi squared $\chi_\mathrm{red}^2$ of the interferogram fit.
All fits with $\chi_\mathrm{red}^2  \geq 1.5$ were discarded and not included in our 2008 results.
This was done to account for phase instabilities especially those seen immediately following a cell transfer which introduced temporary temperature and mechanical instabilities lasting \unit[12]{h} or more.  
However, discarding interferograms based on  $\chi_\mathrm{red}^2$ values included eliminating points taken in the middle of runs where the phase was more stable.
As discussed below, a systematic uncertainty of \unit[0.012]{fm} attributed to phase instabilities was also applied  to the result in Ref.~\cite{Huber_2009_PRLa}.
Since we already incorporate an uncertainty due to phase instabilities, for this result we make no cut based on the  $\chi_\mathrm{red}^2$ in either the 2008 or 2013 data set.
Phase instabilities were greater in 2013 as the temperature stability that we had in 2008 was not reproduced. 
This is contrary to 2008 where  temperature drifts  were highly correlated to  opening the facility  doors to perform a cell transfer (since $P_3 \approx 0$  in most of  2013, transfers were infrequent in that data run).
The inclusion of data points with $\chi_\mathrm{red}^2  \geq 1.5$  does not affect the values determined by a fit of $\Delta\phi_0$ versus $\xi$ but does decrease the statistical uncertainty.
\par
A two parameter fit of Fig.~\ref{Fig_Result_2}a gives $\Delta b^\prime =$  \werb{$-$5.381}{0.053}{fm} with $\chi^2/d.o.f.= 530/(435-2)=1.2$.  
This $\chi^2$ represents a low probability of fit ($<$ \unit[1]{\%}) and is due to random phase instabilities that were most likely caused by small temperature fluctuations. 
To estimate the systematic uncertainty due to this effect the uncertainty of $\Delta\phi_0$ was inflated by \unit[0.016]{rad}  in quadrature with the statistical uncertainty for each point  so that the $\chi^2/d.o.f.=1$. 
The average statistical uncertainty for $\Delta\phi_0$ was  $\approx$ \unit[0.033]{rad} but varied strongly with $P_3$.
A histogram of the residual of the fit with a reduced $\chi^2=1$ is shown in the inset of  Fig.~\ref{Fig_Result_2}a.
The distribution of points in the figure closely follows a Gaussian function centered at zero.  
\par
Figure \ref{Fig_Result_2}b  shows $\Delta\phi_0$ versus $\xi$ for the 2013 data set.  
In 2013 we polarized the \he\ gas only four times focusing instead on measuring $\Delta\phi_0$ at low $P_3$.
Twice we polarized the \he\ in the opposite direction  with respect to the neutron polarization defined by the supermirror polarizer.
In this case  there is more neutron absorption when the  precession coil spin flipper  is off.  
This reversed-polarized data is shown in  the lower left quadrant  of Fig.~\ref{Fig_Result_2}b.
Again applying a two parameter fit of Fig.~\ref{Fig_Result_2}b gives  $\Delta b^\prime =$ \werb{$-$5.439}{0.038}{fm} with a $\chi^2/d.o.f.= 1120/(507-2)=2.2$. 
To fix  $\chi^2/d.o.f.=1$ the uncertainty of $\Delta\phi_0$ was inflated  in quadrature by \unit[0.043]{rad}.
For 2013 we find that $2\phi_\mathrm{mag} =$ \werb{21}{3}{mrad}  which is consistent with both the 2008 data and the field gradient measurement. 
\par
The weighted average of both data sets gives 
\begin{eqnarray}
 \Delta b^\prime &=& \label{Eq_Result_1}\\ 
& &\mbox{\WERB{\ComDb}{\ComDbstat} {Stat.}{\ComDbsys}{Syst.}{fm.}} \nonumber
\end{eqnarray}
This corresponds to a $4\sigma$ shift of $\Delta b^\prime $  compared to our previous result reported in Ref.~\cite{Huber_2009_PRLa}.
This shift is entirely due to the inclusion of phase shifts from magnetic field gradients in our fitting.
Allowing our fit of $\Delta\phi_0/\xi$ an additional degree of freedom increased the statistical uncertainty in the scattering length by a factor of $2$.
However, tripling the original data set yielded a final statistical uncertainty only \unit[20]{\%} larger than that reported in Ref.~\cite{Huber_2009_PRLa}.
The uncertainty budget for $ \Delta b^\prime $  for each individual data set is given in Table \ref{table3}.  
The weighted average is performed by weighting  both the statistical and  systematic uncertainties unrelated to neutron  absorption on \he\  in quadrature. 
The systematic uncertainty related to \he\ absorption was added to the total systematic uncertainty in \eq{Eq_Result_1}.
\begin{table}{\caption{The uncertainty budget for  $\Delta b^\prime$. Uncertainties related to \he\ absorption  cross section that are identical for both data sets  are summed in quadrature.  }\label{table3}}
		\centering  
		\begin{tabular}[c]{l@{\hspace{0.1in}}c@{\hspace{0.1in}}r}
		\hline\hline
\multicolumn{1}{c}{2008}    &       &   \multicolumn{1}{c}{2013}     \\
     $\sigma$, [fm]                    &     Parameter & $\sigma$, [fm]     \\
																		\hline
		  0.053 & $\Delta\phi_0/\xi$ \quad Fit  (Statistical)                                     &  0.038 \\\hline 
			0.028 &Triplet absorption cross section $\sigma_1$                    &  0.028 \\
			0.007 &Total absorption cross section $\sigma_\mathrm{un}$ &  0.007   \\
			0.029 &Total systematic from  cross sections                                & 0.029 \\
			\hline
			0.025  &Phase instabilities                                                                &  0.040\\
			0.005  &Neutron polarization $P_n$                                                & 0.004\\
			0.002  &Spin flipper efficiency $s$                                                   & 0.004\\
			0.026  &Total non-cross section systematic                                  & 0.040 \\
			\hline
			 0.053 & Total statistical                                                                 &  0.038\\  
      0.039 &Total  systematic                                                                 & 0.049\\
 	\hline\hline
		\end{tabular}		
\end{table}

\section{Conclusions and Discussion \label{Sec_Con}}
We have performed a precision measurement of the difference $\Delta b^\prime = $ \WERB{\ComDb}{\ComDbstat} {Stat.}{\ComDbsys}{Syst.}{fm} between the  triplet and singlet scattering lengths of \nhe\ using neutron interferometry to \unit[0.9]{\%} relative standard uncertainty. 
The ultimate precision of this technique  is  systematically limited by the triplet absorption cross section  corresponding to a relative uncertainty of \unit[0.5]{\%}.
This result is in good agreement with the only previous direct measurement of  $\Delta b^\prime = $ \werb{-5.462}{0.046}{fm} performed by Zimmer \etal at the Institut Laue-Langevin (ILL) \cite{Zimmer_2002_EPJ}.  
Ref.~\cite{Zimmer_2002_EPJ} used a spin echo apparatus to measure the relative difference in the pseudomagnetic spin-precession \cite{Baryshevsky_1965_JETP,Abragam_1973_PRL} between a neutron passing though a polarized \he\ cell compared to an empty reference beam. 
That technique is fundamentally different than the technique applied here.  
One can state the results independent of the triplet absorption cross section and total absorption cross section from our results and that of  Ref.~\cite{Zimmer_2002_EPJ}.
This is done for two reasons: (i) both groups estimated  $\sigma_1$  differently and (ii) in the event of future more accurate measurements of the absorption cross sections, one can immediately update the spin-dependent \nhe\ scattering length.
Zimmer \etal  determined $\sigma_1$ from an average of the experimental results of Refs.~\cite{Passell_1966_PhysRev} and \cite{Borzakov_1982_SovJNuclPhys} with the limitation that $\sigma_1 \geq 0$.
Whereas, as described in Section \ref{Sec_Abs}, we used a theoretically predicted  $\sigma_1$ but with an inflated uncertainty. 
Our result  stated independent of the triplet absorption cross section is
\begin{align}
\Delta b^\prime (\mbox{this work})  = \quad\quad& \nonumber \\ 
\bigg[(-10.1929 &\pm 0.0760) \times 10^{-4}\mbox{ fm/b} \bigg]  \nonumber \\ 
\times \bigg(1-&\frac{\sigma_1}{\sigma_\mathrm{un}}\bigg)\sigma_\mathrm{un}
\end{align}
This is in disagreement with the result of Zimmer \etal of  
\begin{align}
\Delta b^\prime  (\mbox{Ref.~\cite{Zimmer_2002_EPJ}})= \quad\quad& \nonumber \\ 
\bigg[(-10.3628 &\pm 0.0180) \times 10^{-4}\mbox{ fm/b} \bigg]  \nonumber \\ 
\times \bigg(1-&\frac{\sigma_1}{\sigma_\mathrm{un}}\bigg)\sigma_\mathrm{un}
\end{align}
by $2\sigma$ when factoring out the absorption cross sections. 
\begin{figure}[t!]
	\centering
	\includegraphics[width=3.375in]{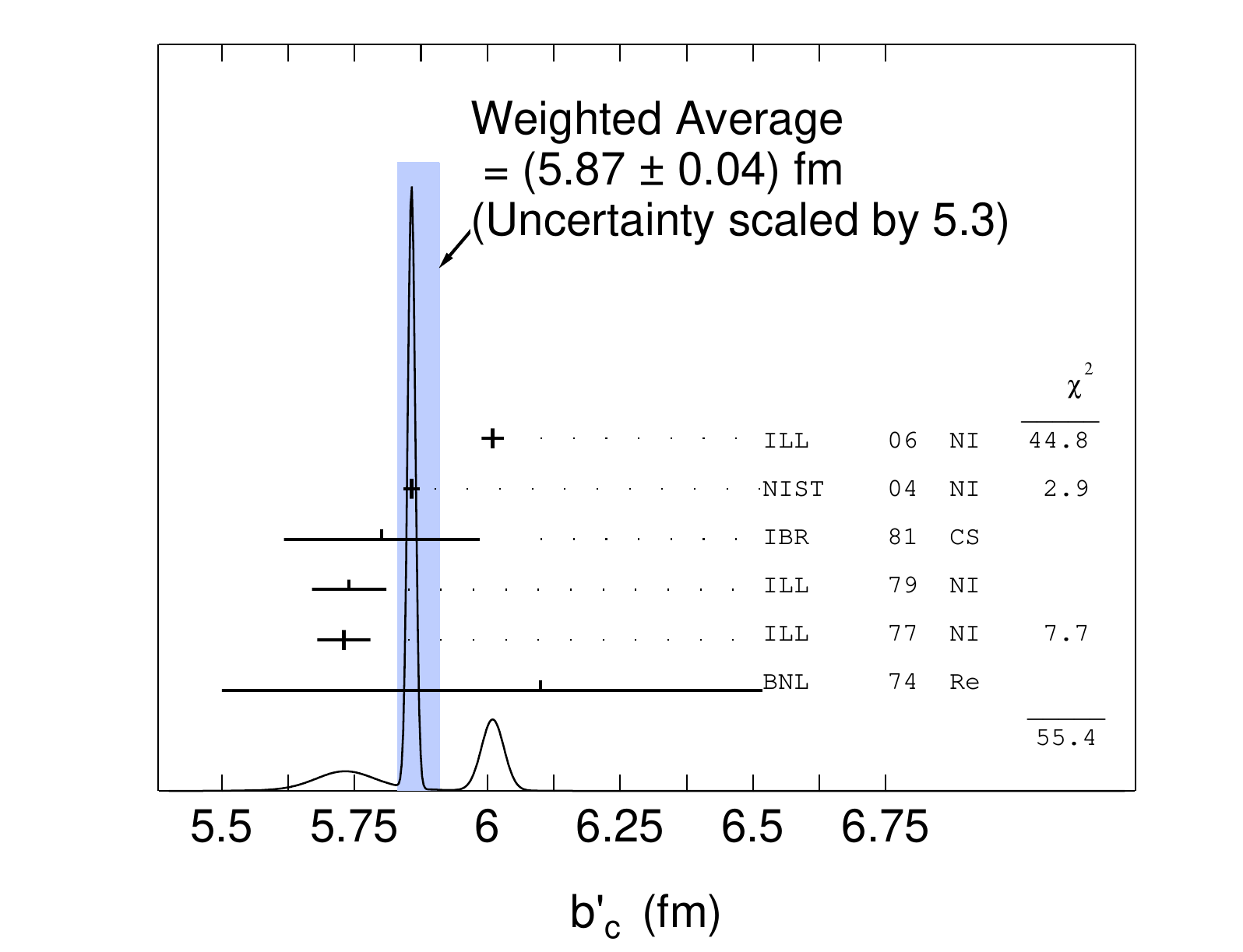}
	\caption{(Color online) An ideogram of the coherent scattering length measurements for \nhe\ taken from Refs.~\cite{Kitchens_1974_PRL,Kaiser_1977_PhysLett,Kaiser_1979_ZPhys,Alfimenkov_1981_SovJNuclPhys,Huffman_2004_PhysRev,Ketter_2006_EPJ}.  The blue band represents the weighted average $\pm\sigma$ of the three experiments with smallest quoted uncertainties. Techniques used were neutron interferometry (NI), total cross section (CS), and reflectivity (RE).}  
	\label{fig13}
\end{figure}
\begin{figure}[b]
	\centering
		\includegraphics[width=3.375in]{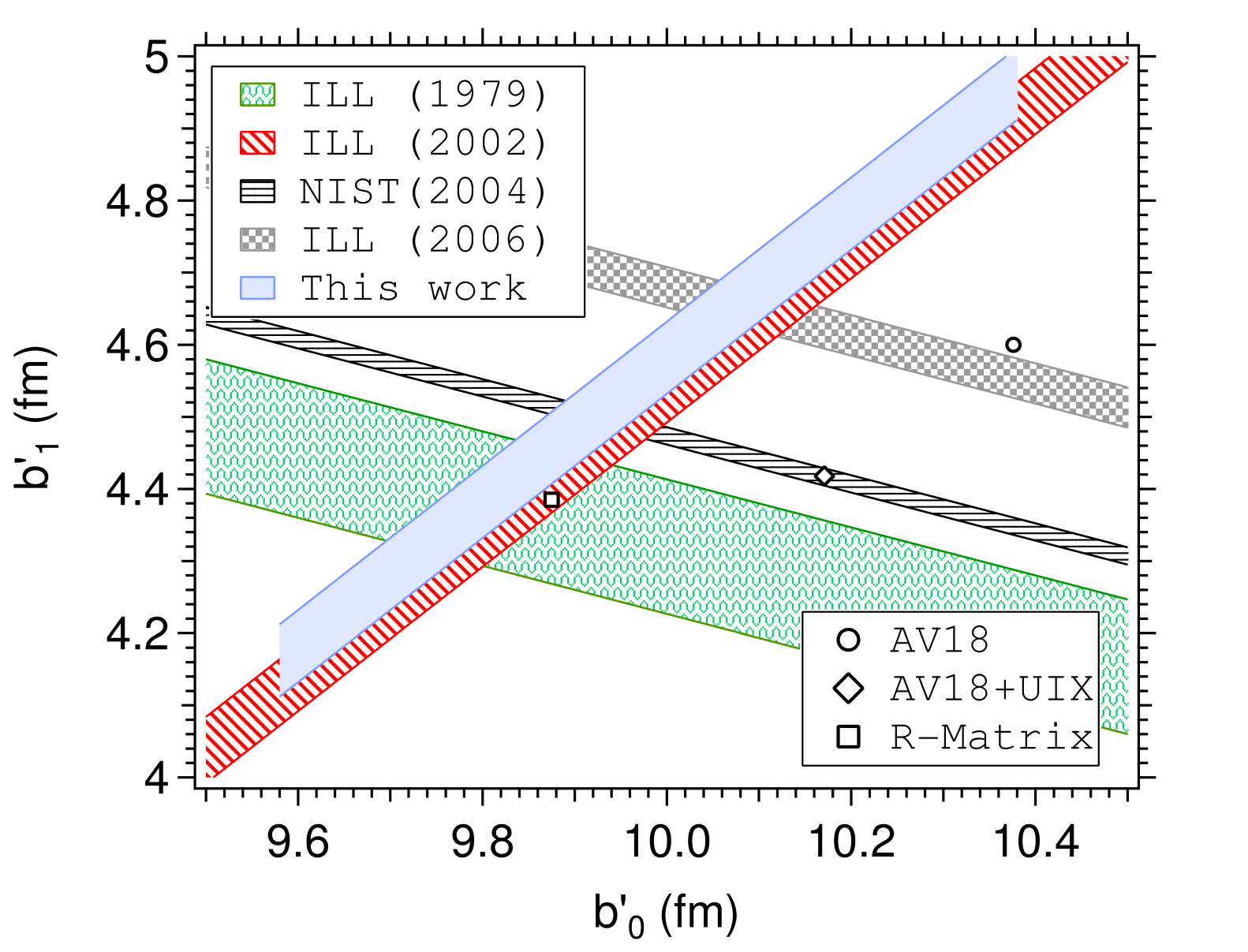}
	\caption{(Color online) Current experimental data on the \nhe\ system from this work, ILL 2006 
	\cite{Ketter_2006_EPJ}, NIST 2004 \cite{Huffman_2004_PhysRev}, ILL 2002 \cite{Zimmer_2002_EPJ}, ILL 1979 \cite{Kaiser_1979_ZPhys} compared to 
	theoretical predictions \cite{Hofmann_2003_PhysRev, Hofmann_2008_PhysRev}.  
	Bands represent the experimentally determined values 
	$\pm 1\sigma$. }  
	\label{fig14}
\end{figure}

\par
There have been a number of experiments measuring the coherent scattering length of \nhe\ defined by \eq{Eq_Theo_11} using  techniques such as measuring neutron reflectivity, relative phase shifts, and neutron transmissions. 
The three most precise measurements of $b_c^\prime$ were done with neutron interferometry;  Kaiser \etal \cite{Kaiser_1977_PhysLett}, Huffman \etal \cite{Huffman_2004_PhysRev}, and Ketter \etal \cite{Ketter_2006_EPJ}. 
However, the two most recent  results differ by more than $7\sigma$.  
Figure \ref{fig13} shows an ideogram of the coherent scattering length measurements. 
Each measurement is represented by a Gaussian centered about their result with a normalized area equal to 1/$\sigma$ \cite{Amsler_2008_PhysLett}.
The uncertainty of the weighted average has been inflated in the manner described in Ref.~\cite{Amsler_2008_PhysLett}.  
\par
Calculations employing models AV18+UIX, AV18+UIX+V$^*_3$ \cite{Hofmann_2003_PhysRev,Hofmann_2008_PhysRev}, and AV18+LL2 \cite{Kirscher_2009_Archive} have all predicted similar values for the triplet and singlet scattering lengths. 
For example $\Delta b^\prime(AV18+UIX) = $ \werb{-5.753}{0.002}{fm}.
Neither this work nor the work of Zimmer \etal agrees with NN+3N calculations.
Figure \ref{fig14} shows a selection of measured values of  $b_1^\prime$ and $b_0^\prime$ beside some theoretical predictions.
Four nucleon interactions have yet to be included into the theoretical models due to the difficulty in handling long-range Coulomb forces, but should  constitute only a tiny correction to NN + 3N predictions.
A calculation of pion-less effective field theory to Next to Leading Order (NLO) shows promise \cite{Kirscher_2009_Archive}, but the uncertainty of the predicted value is still too large to compare to high-precision measurements.
A recent measurement of the total scattering cross section \cite{Guckelsberger_2000_Physica} that suggests a much larger scattering cross section and would lie outside of Fig.~\ref{fig14} is omitted for space.
\par
The recent work on the \nhe\ interaction can lead to further understanding of low energy nucleon systems.  
Although there are several discrepant measurements, scattering length measurements do not match theoretical models.  
Taken alone, the coherent scattering length  by  \cite{Huffman_2004_PhysRev} agrees with AV18+UIX,  but doesn't intersect a measurement of the spin-dependent difference in triplet and singlet states. 
This work and \cite{Zimmer_2002_EPJ}  agrees with the R-matrix prediction. 
More work needs to be done to resolve the discrepancy between different \nhe\ coherent scattering length measurements. 
The uncertainty in the triplet absorption cross section needs to be  experimentally determined to better precision, if other measurements of the spin-dependent quantity $\Delta b^\prime$ are to be made. 
The authors hope that this work along with the previous scattering length measurements can improve future NN+3N models, and is part of the ongoing exploration into few-body systems at the NIOF.

\begin{acknowledgments}
We wish to thank John Fuller and Jeff Anderson at NIST for making the glass target cells. 
The development and application of the polarized \he\ cells and methods used in this experiment was supported in part by the Department of Energy, Basic Energy Sciences. 
Also we would like thank Sam Werner and Helmut Kaiser for their helpful discussions.   
This work is supported by NIST and the National Science Foundation through grants PHY-0555347, PHY-0855445, and PHY-1205342.
\end{acknowledgments}

\bibliographystyle{apsrev4-2}  
\bibliography{nHe2013_v2p1}


\clearpage
\newpage

\begin{center}
	\textbf{\large A Comment on the Helium-3 Triplet Absorption Cross Section:}\\ Comparing neutron interferometry with that of a previous spin echo result\\[.2cm]
	Michael G. Huber,$^{1,*}$ Thomas R. Gentile,$^{1}$ \\[.1cm]
	{\itshape ${}^1$National Institute of Standards and Technology\\}
	${}^*$michael.huber@nist.gov\\
	(Dated: 3/10/2023)\\[1cm]
\end{center}

\setcounter{equation}{0}
\setcounter{figure}{0}
\setcounter{table}{0}
\setcounter{page}{1}
\renewcommand{\theequation}{S\arabic{equation}}
\renewcommand{\thefigure}{S\arabic{figure}}

As of 2022, there are only  two groups that have measured the incoherent (spin-dependent) scattering length of $^3$He. The earliest of which utilized a spin echo apparatus and reported a result in Ref.~\cite{Zimmer_2002_EPJ} (Zimmer). The second group reported their result here and in Ref.~\cite{Huber2014} (Huber).  Although differences in the two results are covered in the preceding archive draft and the subsequent Physics Review C article, there are, as always, some points to clarify.   This is document's  main purpose is connect the subtle differences in notation used by both groups and is not necessary to understand work as a whole.  For this document I'll use the notation of Huber et al. \cite{Huber2014} through out; making changes to Zimmer's text when appropriate. A table comparing the two is given below along with a listing of various values used by both groups. 

\section{A comment on Zimmer et al. result}
Zimmer et al. \cite{Zimmer_2002_EPJ} reported (Zim-Eqn. 55) a result of 
\begin{eqnarray}
	b^\prime_i = -2.365 \quad \mathrm{fm} \nonumber
\end{eqnarray}
for the incoherent scattering length of $^3$He. (For this supplementary post, I will forego any uncertainties).     Ref. \cite{Zimmer_2002_EPJ} also reported their result independently of the triplet absorption cross section $\sigma_1$ as (Zim-Eqn. 52)
\begin{eqnarray}
	b^\prime_i = -0.8068 \times \bigg( \frac{4x_- -1}{3}\bigg)K_a  \nonumber
\end{eqnarray}
For a more direct comparison between the two groups one can convert this result into the scattering length difference. Since 
\begin{eqnarray}
	\Delta b^\prime &=& b^\prime_1 - b^\prime_0 = \frac{b^\prime_i}{\sqrt{g_1g_0}} \quad \quad\mathrm{(see \  Zim-Eqn. 7)} \nonumber
\end{eqnarray}
we have 
\begin{eqnarray}
	\Delta b^\prime &=& \frac{4}{\sqrt{3}} \times (-0.8068)   \times \bigg( \frac{4x_- -1}{3}\bigg)K_a. \nonumber
\end{eqnarray}
One can go a step further and convert $K_a$ into an absorption cross section since $K_a =\sigma_{un}/\lambda_{th}$ (Zim-Eqn. 22).  Then
\begin{eqnarray}
	\Delta b^\prime &=& \frac{4}{\sqrt{3}} \times (-0.8068) \times \bigg( \frac{4x_- -1}{3}\bigg)\frac{\sigma_{un}}{\lambda_{th}} \\ \nonumber
	&=& [ -10.3628\times 10^{-4} \quad \mathrm{ fm/b} ]\times\bigg( \frac{4x_- -1}{3}\bigg)\sigma_{un}
\end{eqnarray}
\par
Now let us see how the triplet absorption cross section can be isolated in the form that was reported by Ref. \cite{Huber2014}. Specifically, in terms of 
\begin{eqnarray}
	\left( 1- \frac{\sigma_{1}}{\sigma_{un}}\right). \label{HubRatio}
\end{eqnarray}
We will start with the observation that
\begin{eqnarray}
	\sigma_{un} &=& g_1\sigma_1+g_0\sigma_0 \quad \mathrm{(by  \ defn.)}. \label{eqn3}
\end{eqnarray}
Zimmer defines the ratio $x_-$ as
\begin{eqnarray}
	x_-=\frac{g_0\sigma_{0}}{\sigma_{un}} \label{xxx} 
\end{eqnarray}
Rearranging Eqn. \ref{xxx} we can write 
\begin{eqnarray}
	g_0\sigma_{0} &=& x_-\sigma_{un} .
\end{eqnarray}
Combined with Eqn. \ref{eqn3} we have
\begin{eqnarray}
	\sigma_{un} &=& g_1\sigma_1+x_-\sigma_{un}   
\end{eqnarray}
or alternatively, 
\begin{eqnarray}
	g_1\sigma_1 &=& (1-x_-)\sigma_{un}. \label{sigUN}
\end{eqnarray}
We can plug Eqn. \ref{sigUN} into Eqn. \ref{HubRatio}. 
\begin{eqnarray}
	\left( 1- \frac{\sigma_1}{\sigma_{un}}\right) &=&\left[1 -\frac{(1-x_-)\cancel{\sigma_{un}}}{g_1\cancel{\sigma_{un}}} \right]\nonumber \\
	&=&\left[ 1  -\frac{4}{3}(1-x_-)  \right] \nonumber\\
	&=&\left( 1  -\frac{4}{3} + \frac{4}{3}x_-  \right) \nonumber\\
	&=&\left(  -\frac{1}{3} + \frac{4}{3}x_-  \right) \nonumber\\
	&=&\left( \frac{4x_--1}{3}  \right) \nonumber\\
	\left( 1- \frac{\sigma_1}{\sigma_{un}}\right)&=&\left( \frac{4x_--1}{3}  \right) \nonumber
\end{eqnarray}
So now we have shown the equivalence of the form used by Zimmer et al. with that of the form used in Huber et al.

\section{A comment on Huber et al. result}
We will start with the relevant equations from Ref. \cite{Huber2014} which are (Hub-Eqn. 18): 
\begin{eqnarray}
	b_1^\prime -b_0^\prime =\frac{\displaystyle -2\Delta\phi_0}{N_{3}\lambda D_{3} P_{3}}, \label{Eq_Proc_1} 
\end{eqnarray}
and (Hub-Eqn. 61):
\begin{eqnarray}
	\xi=N_{3}\sigma_{p} D_{3}P_3=N_{3}\left[\frac{1}{4}\left(\sigma_0-\sigma_1 \right)\right]\frac{\lambda}{\lambda_\mathrm{th}}D_{3}P_3, \label{Eq_Syst_1}
\end{eqnarray}
Rearranging the second of these we have,
\begin{eqnarray}
	N_{3}\lambda D_{3}P_3= \frac{\xi\lambda_{th}}{\frac{1}{4}(\sigma_0-\sigma_1)} 
\end{eqnarray}
which we can plug into Eqn. \ref{Eq_Proc_1} to get
\begin{eqnarray}
	\Delta b^\prime  &=& b_1^\prime -b_0^\prime \nonumber \\
	&=&\left(\frac{\displaystyle -2\Delta\phi_0}{\xi\lambda_{th}}\right)\left[\frac{1}{4}\left(\sigma_0-\sigma_1\right)\right] \label{esss}
\end{eqnarray}
The variables in the $()$ are simply measured in the experiment and give rise to the factor of -10.1929 in the result of Ref. \cite{Huber2014}.  Namely,  
\begin{eqnarray}
	\left[-10.1929 \times 10^{-4}\mbox{ fm/b} \right]\times
	\left(1-\frac{\sigma_1}{\sigma_\mathrm{un}}\right)\sigma_\mathrm{un}.  \label{HubResult} 
\end{eqnarray}
The $[]$ term in Eqn. \ref{esss} we will manipulate here. Starting with
\begin{eqnarray}
	\left[\frac{1}{4}\left(\sigma_0-\sigma_1\right)\right] &=&  g_0\sigma_0- g_0\sigma_1 \label{eqns10} \\
\end{eqnarray}
since $g_0 = 1/4$.
The trick is to add $0=+g_1\sigma_1-g_1\sigma_1$ to right side of Eq.\ref{eqns10}
\begin{eqnarray}
	g_0\sigma_0- g_0\sigma_1 &=& (g_0\sigma_0- g_0\sigma_1) + (0)  \nonumber\\
	&=& (g_0\sigma_0- g_0\sigma_1) + (g_1\sigma_1-g_1\sigma_1)   \nonumber
\end{eqnarray}
and then rearrange terms
\begin{eqnarray}
	g_0\sigma_0- g_0\sigma_1 + g_1\sigma_1-g_1\sigma_1 &=& (g_0\sigma_0 + g_1\sigma_1) - (g_0\sigma_1+g_1\sigma_1)   \nonumber\\
	&=& \sigma_{un} - (g_0+g_1)\sigma_1  \nonumber
\end{eqnarray}
but $g_0 + g_1 =1$  so 
\begin{eqnarray}
	\sigma_{un} - (g_0+g_1)\sigma_1  &=& \sigma_{un} - \sigma_1  \nonumber\\
	&=& \left(1 - \frac{\sigma_1}{\sigma_{un}} \right)\sigma_{un}  \nonumber
\end{eqnarray}
ending up with the factor in Eqn. \ref{HubResult}.

\pagebreak

\begin{table*}[!htbp]
	\begin{center}
		\begin{tabular}{ l c r }
			Ref. \cite{Zimmer_2002_EPJ}  & Ref. \cite{Huber2014} &   \\
			Notation & Notation &Description \\
			\hline\hline
			$b^\prime_i$ &  $b^\prime_i$ & real part of the incoherent scattering length   \\ 
			$b_+$ & $b_1$ & triplet scattering length    \\
			$b_-$ & $b_0$ & singlet scattering length   \\
			$g_+ =3/4$ & $g_1$ & triplet state weight factor    \\
			$g_- =1/4$ & $g_0$ & singlet state weight factor   \\
			$K_a$ & $\sigma_{un}/\lambda_{th}$ & absorption cross section proportionality constant\\
			$x_- = \frac{g_-\sigma_{a,-}}{\sigma_{a,0}}$ & N/A & relative contribution \\
			$\lambda_{th}$ &$\lambda_{th}$ & thermal neutron wavelength \\
			$\sigma_{a,0}$ & $\sigma_{un}$ & unpolarized absorption cross section \\
			$\sigma_{a,p}$  & $\sigma_{p}$ & polarized  absorption cross section \\
			\hline
		\end{tabular}
		\caption{A short description and comparison of variables used by Zimmer er al. and Huber et al. }
	\end{center}
\end{table*}

\vspace{.25 in}
\begin{table*}[!htbp]
	\begin{center}
		\begin{tabular}{ l c r }
			Symbol & Ref. \cite{Zimmer_2002_EPJ} Value & Source \\ 
			\hline\hline
			$ \sigma_{un}$ & 5327 b & Ref. \cite{AlsNielsen1964} \\
			$ K_a$ & 2.9626  fm & Ref. \cite{AlsNielsen1964} \\ 
			$ \lambda_{th}$ & 0.1798 nm & by defn. \\
			$x_-$ & 0.992 & Refs. \cite{Passell_1966_PhysRev,Borzakov_1982_SovJNuclPhys} with constraint of $x_-\leq 1$ \\
			$ \sigma_{1}$ & 57 b & based on $x_-$ and $ \sigma_{un}$ using Eqn. \ref{esss}
			\\\hline
		\end{tabular}
		\caption{Variables used by Zimmer er al. }
	\end{center}
\end{table*}
\vspace{.25 in}
\begin{table*}[!htbp]
	\begin{center}
		\begin{tabular}{ l c r }
			Symbol & Ref. \cite{Huber2014} Value & Source \\ 
			\hline\hline
			$ \sigma_{un}$  & 5333 b & Ref. \cite{Sears1992} \\
			$ \lambda_{th}$ & 0.1798 nm & by defn.\\
			$ \sigma_{1}$ & 24 b & based on a theoretical calculation of Ref. \cite{Hofmann_2003_PhysRev, Hofmann_2008_PhysRev} \\\hline
		\end{tabular}
		\caption{Variables used by Huber er al. }
	\end{center}
\end{table*}

\end{document}